\documentclass[10pt, conference, compsocconf]{IEEEtran}
\pdfoutput=1

\usepackage{graphicx} 
\usepackage{float}
\usepackage{caption}
\usepackage{subcaption}
\usepackage{chngcntr}
\usepackage{amsmath}
\usepackage{algorithm}
\usepackage{algorithmic}
\usepackage{hyperref}

\begin{document}
%
\title{Towards personalized human AI interaction - adapting the behavior of AI agents using neural signatures of subjective interest}


\author{\IEEEauthorblockN{Victor Shih\IEEEauthorrefmark{1},
David C Jangraw\IEEEauthorrefmark{3},
Paul Sajda\IEEEauthorrefmark{1}\IEEEauthorrefmark{2}, and
Sameer Saproo\IEEEauthorrefmark{1}}
\IEEEauthorblockA{\IEEEauthorrefmark{1}Department of Biomedical Engineering, Columbia University, New York, NY 10027 USA} 
\IEEEauthorblockA{\IEEEauthorrefmark{2}Data Science Institute, Columbia University, New York, NY 10027 USA} 
\IEEEauthorblockA{\IEEEauthorrefmark{3}National Institutes of Health, Bethesda, MD 20892 USA}
Email: vs2481@columbia.edu}

%



\IEEEtitleabstractindextext{%
\begin{abstract}
	Reinforcement Learning AI commonly uses reward/penalty signals that are objective and explicit in an environment -- e.g. game score, completion time, etc. -- in order to learn the optimal strategy for task performance. However, Human-AI interaction for such AI agents should include additional reinforcement that is implicit and subjective -- e.g. human preferences for certain AI behavior -- in order to adapt the AI behavior to idiosyncratic human preferences. Such adaptations would mirror naturally occurring processes that increase trust and comfort during social interactions. Here, we show how a hybrid brain-computer-interface (hBCI), which detects an individual's level of interest in objects/events in a virtual environment, can be used to adapt the behavior of a Deep Reinforcement Learning AI agent that is controlling a virtual autonomous vehicle.  Specifically, we show that the AI learns a driving strategy that maintains a safe distance from a lead vehicle, and most novelly, preferentially slows the vehicle when the human passengers of the vehicle encounter objects of interest.  This adaptation affords an additional 20\% viewing time for subjectively interesting objects. This is the first demonstration of how an hBCI can be used to provide implicit reinforcement to an AI agent in a way that incorporates user preferences into the control system. 
\end{abstract}
}

\maketitle

\IEEEdisplaynontitleabstractindextext

%
\IEEEpeerreviewmaketitle

\section{Introduction}
%
%
%
%
\IEEEPARstart{T}{he} use of Artificial Neural Networks (ANNs) towards developing Artificial Intelligence (AI) has undergone a renaissance in the past decade. Out of the many emergent techniques for training ANNs that are collectively referred to as 'Deep Learning', Deep Reinforcement Learning (DRL) is proving to be a particularly general and powerful method, with applications ranging from video games \cite{mnih:dqn} to autonomous driving \cite{lange:autonomous}. While most applications of reinforcement learning have traditionally used reinforcement signals derived from performance measures that are explicit to the task -- e.g. the score in a game or grammatical errors in a  translation, when considering AI systems that are required to have a significant interaction with humans -- e.g. the autonomous vehicle -- it is critical to consider how the human's preference for objects, events, or actions can be incorporated into the behavioral reinforcement for the AI, particularly in ways that are minimally obtrusive \cite{saproo:cortically,lake:building}. Such behavioral adaptations occur naturally during social interactions and form the bedrock of social mechanisms that build trust and rapport between strangers \cite{ iacoboni:mirroring,hasson:mirroring}.
 
 \begin{figure*}[!h]
 	\centering
 	\includegraphics[width=0.9\textwidth]{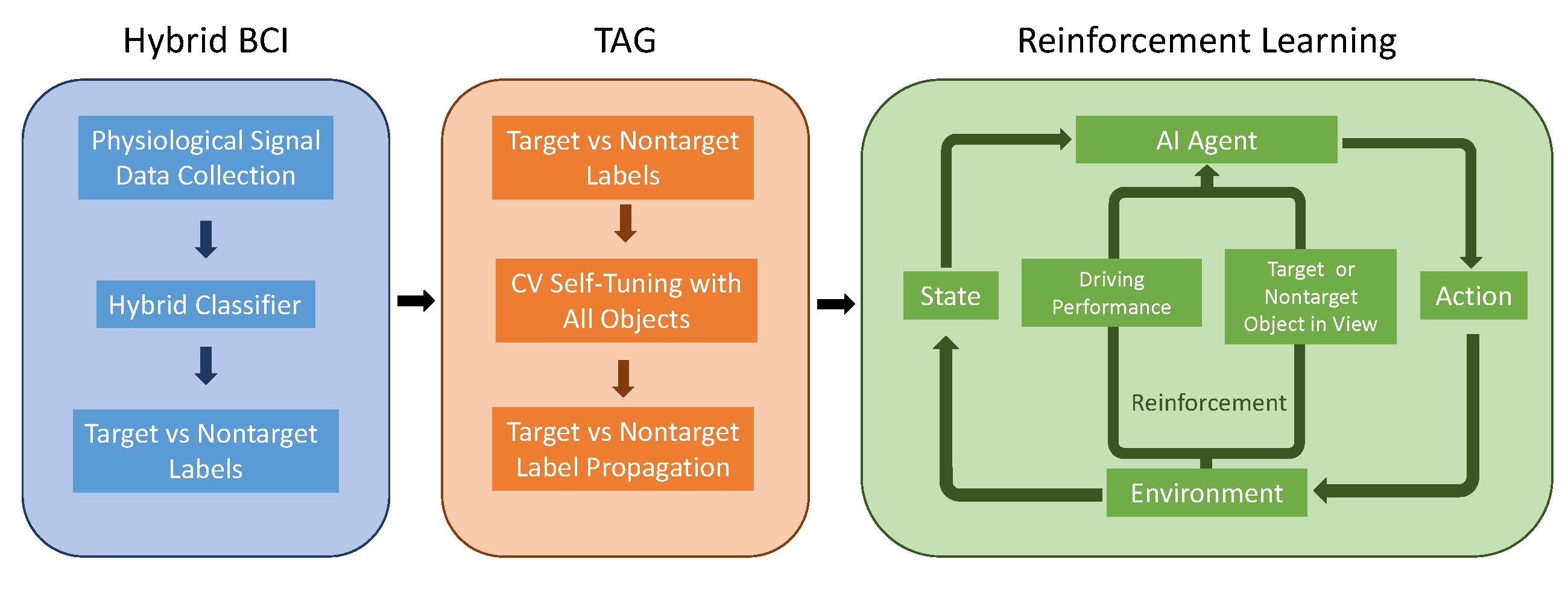}
 	\caption[Experimental Design]%
 	{{\small Machine Learning Schema. The three stages of machine learning that are used in our system. The first stage uses a hybrid classifier to fuse and decode physiological signals from the subject and identify 'target' (interesting) or 'non-target' (uninteresting) objects from a subset of the object database. The second stage uses these object labels and a CV system to identify targets and non-targets from the full object database. The third stage populates a virtual environment with objects and uses the object labels ('target' or 'non-target') to reinforce the AI agent, such that the AI agent learns to keep the target objects in view for longer periods of time.}}    
 	\label{fig:ml_fig}
 \end{figure*}

 In this paper, we  present a novel approach that uses decoded human neurophysiological and ocular time-series data as an implicit reinforcement signal for an AI agent that is driving a virtual automobile. The agent  learns a brake and accelerate strategy that integrates road safety with the personal preferences of the human passenger. These preferences are derived from the neural (EEG: electroencephalography) and ocular signals (pupillometry and gaze time) that are evoked by interesting objects/events in the simulated environment. We integrate and decode these signals and construct a hybrid brain-computer interface (hBCI) \cite{jangraw:3dsearch} whose output represents a passenger's subjective level of interest in objects/events in the world, and therefore can be used to reinforce AI behavior.

 We describe the details of our approach, including the cognitive neuroscience basis of the signals we decode and integrate within the hBCI. We then show how the hBCI can be made more robust by adding a semi-supervised graph-based model of the objects called TAG: Transductive Annotation by Graph \cite{wang:TAG}. This graph-based model reduces errors that may result from the neural-ocular decoding as well as extrapolates the preference estimates derived from a small number of viewed objects to a much larger number of previously unencountered  objects. This extrapolation reduces the amount of neural data required for the system to function effectively. We show that the AI converges to a driving behavior that increases the time that a passenger gets to view objects of interest in the environment. Finally, we discuss the extension of this approach to various human-AI-interaction scenarios that incorporate other measures of an individual's cognitive state, for example, tailoring the experience in an autonomous vehicle based on their level of comfort or arousal.	

\section{Methods}
We used a three-stage machine learning structure to 1) capture neural and physiological signatures of subject preferences, 2) tune and extrapolate these preferences over objects in a given environment, and 3) reinforce driving behavior using a deep reinforcement network (see Figure~\ref{fig:ml_fig}). Specifically, we utilized physiologically derived information from an hBCI to infer subject interest in objects in the environment, classified the objects as targets or non-targets, and used these neural markers in TAG semi-supervised learning architecture \cite{wang:TAG} to extrapolate from a small set of target examples so as to categorize a large database of objects into targets and non-targets. We then placed these target and non-target objects in a virtual environment where the AI drove a simulated automobile. When these objects were in view of the vehicle, we sent positive or negative reinforcement to the learning agent. Our goal was to differentiate the behavior of the AI when driving near targets and non-targets. We incentivized the AI agent to keep within visual distance of targets for a longer period of time.


\begin{figure}[!]
	\centering
	\begin{subfigure}[b]{0.4\textwidth}
		\centering
		\includegraphics[width=\textwidth]{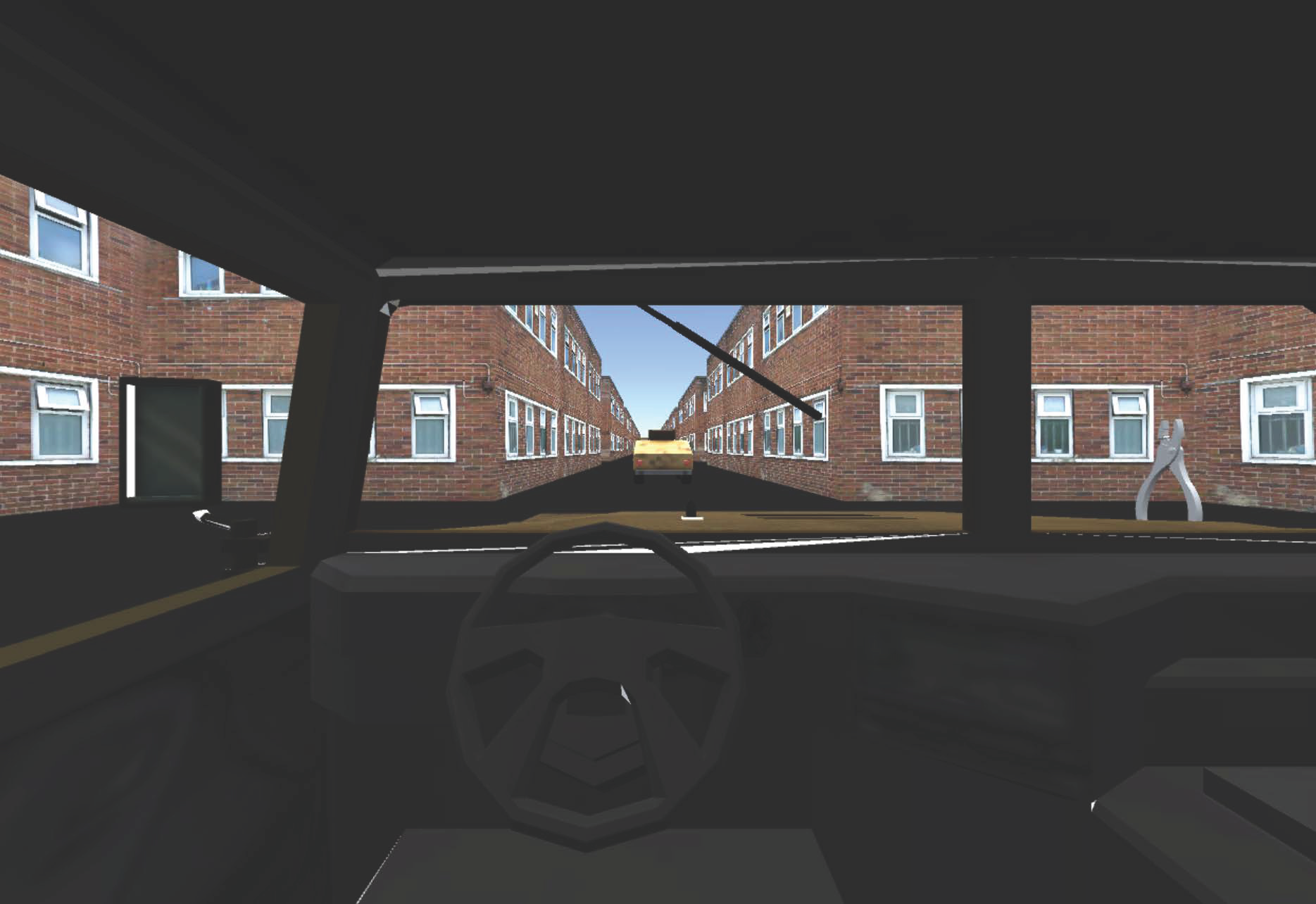}
		\caption[subject view]%
		{{\small Subject View}}    
		\label{fig:virtualenvironment}
	\end{subfigure}
	\begin{subfigure}[b]{0.4\textwidth}
		\centering
		\includegraphics[width=\textwidth]{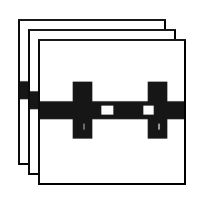}
		\caption[input to deepq]%
		{{\small Input to DeepQ}}    
		\label{fig:rlinput}
	\end{subfigure}
	\caption
	{{\small Virtual Environment for the Experiment. a) A screen capture of the passenger's view in the virtual environment. Objects can be seen in the alleys to the side of the vehicle. b) The input to the AI that shows a top-down perspective of the virtual environment. The two vehicles - the lead vehicle and the AI-controlled vehicle - are seen on the road as white blocks while objects in the alleys are represented by markers with one luminance corresponding to targets and one luminance corresponding to non-targets. This view is intended to mirror the composite maps currently used to train self driving cars \cite{madrigal:waymo}.}}
	\label{fig:experiment}
\end{figure}
\subsubsection{Virtual Environment}
We designed a virtual environment in Unity3D and used assets from a previous study to create objects seen in the virtual environment (Figure \ref{fig:virtualenvironment}) \cite{jangraw:3dsearch}. The vehicle in our simulation had access to a top-down view of the environment that outlines the road, the car in front, and the objects by the side of the road (Figure \ref{fig:rlinput}). We chose this environment representation due to the fact that some autonomous vehicles under development currently are employing a similar composite maps of the environment\cite{madrigal:waymo}. In our virtual environment, a passenger-bearing vehicle controlled by the AI drove behind a pre-programmed lead vehicle that followed a straight path but braked and accelerated stochastically. Each independent driving run in the experiment started from the same location in the virtual environment and would end immediately if the passenger vehicle lagged too far behind the lead vehicle ($>$60m) or followed dangerously close ($<$5m). The AI agent had 3 actions at its disposal: increase speed, maintain speed, or decrease speed of the passenger car. There were alleys on either side of the road, with 40\% of these alleys containing distinctly visible objects. Targets and non-targets objects were placed randomly in the alleyways in a 1:3 prevalence ratio.

\subsection{Subject Preferences}
We tracked the subjective preferences of the human passenger through decoded physiological signals of the human orienting response \cite{nieuwenhuis:orienting}. Orienting is critical to decision-making since it is believed to be important for allocating attention and additional resources, such as memory, to specific objects or events in the environment. Salience and emotional valence are known to affect the intensity of the orienting response.  Orienting is expressed neurophysiologically as evoked EEG, specifically in the P300 response \cite{donchin:p300}. It is also often associated with changes in arousal level, seen in the dilation of the pupil, as well as changes in behavior, for example physically orienting to the object or event of interest \cite{nieuwenhuis:orienting}.

Reinforcement learning typically requires a large number of samples to train a network to reach a successful model. In this experiment, we used physiological signals as reinforcement, but due to the large amount of training data needed, it was unreasonable to have a subject sit through the experiment for the many hours needed to train the network. Instead of using real time physiological reinforcement to train the network, we utilized target object labels from a previous experiment derived from neurophysiological data of subjects and extrapolated to the full object database using the TAG computer vision system \cite{jangraw:3dsearch}. In this way, we were able to build a model that predicted subject preferences but expanded the training dataset so that an accurate AI model could be trained in the virtual environment. 

\begin{figure*}[!h]
	\centering
	\includegraphics[width=0.74\textwidth]{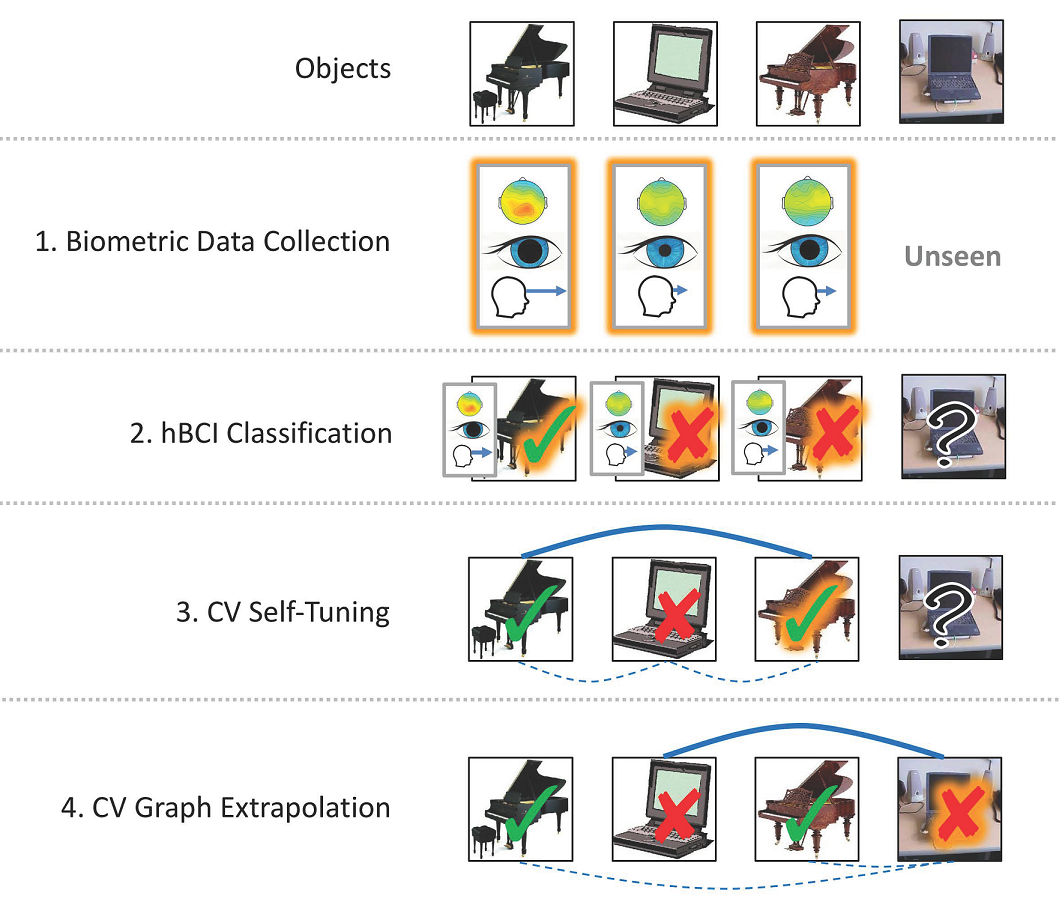}
	\caption[input to deepq]%
	{{\small Illustrative toy example of hBCI + CV labeling of objects in the 3D environment. Top row: objects were placed in the environment, and subjects were asked to count their "preferred" target category (in this example, grand pianos) as they moved through the environment.  Rows numbered Steps 1-4: process by which the objects were labeled as targets (green checks) or non-targets (red x's). The labels could then be used to determine the reinforcement signal sent to the DL system: having a target-labeled object in view resulted in an increased reward. Orange outlines indicate that the data/label was generated in this step. Step 1: The subject viewed some (but not all) of the objects. EEG, pupil dilation, and dwell time data were collected as the subject viewed each one, as described in \cite{jangraw:3dsearch}. Step 2: A subject-specific hBCI classifier was constructed to convert these biometric signals into a target/non-target label. Step 3: the TAG CV system \cite{wang:TAG} was used to "self-tune" the labels, adjusting them so that the predicted targets are strongly connected to each other but not to the predicted non-targets. Blue lines show the level of connection in the CV graph: Thick solid lines represent strong connections; thin dotted lines represent weak ones. Step 4: The tuned labels are propagated through the CV graph to generate labels for the unseen objects.}}    
	\label{fig:hbci}
\end{figure*}
\subsubsection{Hybrid BCI}

In a previous study \cite{jangraw:3dsearch}, subjects were driven through a grid of streets and asked to count image objects of a pre-determined target category. The physiological signals that were naturally evoked by objects in this task were classified by an hBCI system (Fig.~\ref{fig:ml_fig}). This hBCI system was adapted from the hierarchical discriminant component analysis (HDCA) described in Gerson et al (2006), Pohlmeyer et al (2011) and Sajda et al (2010) to accommodate multiple modalities: EEG, pupil dilation, and gaze time. To construct the classifier for each subject, EEG data in the 100 ms to 1000 ms window after the subject fixated on the object were divided into nine 100 ms bins. Within-bin weights across the Independent Components of the EEG data were determined for each bin using Fisher linear discriminant analysis (FLDA):
\begin{equation}
w_j=(\sum_{+}+\sum_{-})^{-1}(\mu_+-\mu_-)
\end{equation}
where $w_j$ is the vector of within-bin weights for bin $j$, $\mu$ and $\Sigma$ are the mean and covariance of the data (across training trials) in the current bin, and the + and − subscripts denote target and non-target trials, respectively. The weights $w$ were applied to the IC activations $x$ from a separate set of evaluation trials to get one ‘within-bin interest score’ $z_{ji}$ for each bin $j$ in each trial $i$ so that:
\begin{equation}
z_{ji}=w^T_jx_{ji}
\end{equation}
The within-bin interest scores from the evaluation trials served as part of the input to a cross-bin classifier. The use of an evaluation set ensured that if the within-bin classifier over-fitted to the training data, this over-fitting would not bias the cross-bin classifier towards favoring these features. 

The pupil dilation data from 0 to 3000 ms were separated into six 500-ms bins and averaged within each bin. For each bin, this average was passed through FLDA to create a discriminant value. The gaze time data was also passed through FLDA. The scale of each EEG, pupil dilation and gaze time feature was then rescaled by dividing each feature's output by its standard deviation across all evaluation trials. A second-level feature vector $z_i$ was created for each evaluation trial $i$ by appending that trial’s rescaled EEG, pupil dilation, and dwell time features into a single column vector. 

To classify the second-level feature vectors from each trial ($z_i$), ‘cross-bin’ weights $v$ were derived using logistic regression, which maximizes the conditional log likelihood of the correct class:
\begin{equation}
v = \arg\min_v(\sum_{i}\log{1+\exp[-c_iv^Tz_i]}+\lambda||v||^2_2)
\end{equation}
where $c_i$ is the class (+1 for targets and −1 for non-targets) of trial $i$ and $\lambda = 10$ is a regularization parameter introduced to discourage over-fitting. These weights were applied to the within-bin interest scores from a separate set of testing trials to get a single ‘cross-bin interest score’ $y_i$ for each trial:
\begin{equation}
y_i=v^Tz_i
\end{equation}
The effectiveness of the classifier was evaluated by its ability to produce cross-bin interest scores $y_i$ that are higher for targets than for non-targets. Trials with cross-bin interest scores more than 1 standard deviation above the mean were identified as ‘hBCI predicted targets’.

\subsubsection{Transductive Annontation by Graph}
TAG used a graph-based system to identify target objects that are of interest to the subject. TAG first tuned the target set predicted by the hBCI for each subject and then extrapolated these results to all the unseen objects in the environment \cite{wang:TAG,jangraw:3dsearch}. TAG constructed a ‘CV graph’ containing all the objects in the virtual environment, using their similarity to determine connection strength (Wang et al 2008, 2009a). The graph employed ‘gist’ features (low-dimensional spectral representations of the image based on spatial envelope properties, as described in Oliva and Torralba (2001)). The similarity estimate for each pair of objects was based not only on the features of that pair, but also on the distribution of features across all objects represented in the CV graph. TAG tuned the hBCI predicted target set by removing objects that did not resemble the set as a whole and replacing them with images that did (Sajda et al 2010, Wang et al 2009a, 2009b). Conceptually, the objects in the hBCI predicted target set that were least connected to the others were deemed most likely to be false positives. They were removed from the set and replaced with the objects not in the set that were most connected to the set. Objects in the resulting set were called ‘tuned predicted targets’

The tuned predicted target set was propagated through the CV graph to determine a ‘CV score’ for each object in the virtual environment, such that the images with the strongest connections to the tuned predicted target set were scored most highly. A cutoff was determined by fitting a mixture of two Gaussians to the distribution of CV scores and finding the intersection point of the Gaussians that falls between their means. The images with CV scores above the cutoff were identified as ‘CV predicted targets’. Because each object was paired with an object in virtual environment space, these CV predicted targets represent the system’s predictions of the objects in the environment that are most likely considered targets by the subject.

\subsection{Deep Reinforcement Learning}
To train the AI agent to navigate the virtual environment, we used a deep reinforcement learning paradigm \cite{mnih:dqn} that optimizes the function for learning the correct action under a given state $S$ using the equation:
\begin{equation}
Q_\pi(s,a) = \mathbf{E}[R_1+\gamma R_2 + ... | S_0=s, A_0=a,\pi]
\end{equation}
Where $\mathbf{E}[R_1]$ is the expected reward of the next state and action pair, and subsequent state action pairs are discounted by $\gamma$ compounding. This Q-function can be approximated by the parameterized value: $Q(s,a;\theta_t)$. Where $\theta_t$ is the parameterized representation of $\pi$. By utilizing reinforcement learning the network builds a model that predicts future states and future rewards in order to optimally accomplish a task. We implemented double-deepQ learning \cite{vanHasslet:ddqn} to update the network weights to this parameterized function after taking action $A_t$ at state $S_t$ and observing the immediate reward $R_{t+1}$ and state $S_{t+1}$ using the equation:
\begin{equation}
\theta_{t+1}=\theta_t+\alpha(Y_t^{DQ}-Q(S_t,A_t;\theta_t))\nabla{\theta_t}Q(S_t,A_t;\theta_t)
\end{equation}
where $\alpha$ is a scalar step size and the target $Y_t^{DQ}$ is defined as:
\begin{equation}
Y_t^{DQ}=R_{t+1}+\gamma Q(S_{t+1},\arg\max_aQ(S_{t+1},a;\theta_{td});\theta_{td})
\end{equation}
By implementing this form of learning, we were able to adjust the reward value to combine explicit reinforcement of driving performance with physiologically derived feedback to influence the behavior of the AI-controlled virtual car.

\subsubsection{Network Architecture}
The deep network used to parameterize the Q-function was created using a 5 layer deep network \cite{mnih:dqn}. We used convolution layers in the network so as to allow computer vision capabilities that can interpret the input state image and identify objects in the image such as the car position and object positions.The input to the neural network consisted of the 3x64x64 grayscale image series state input. The first hidden layer convolved 32 filters of 8x8 with stride 4 with the input image and applied a rectifier nonlinearity. The second hidden layer convolved 64 filters of 4x4 with stride 2, again followed by a rectifier nonlinearity. This was followed by a third convolution layer that convolved 64 filters of 3x3 with stride 1 followed by a rectifier. The final hidden layer was fully-connected and consisted of 512 rectifier units. The output layer was a fully-connected linear layer with a single output for each valid action - increase speed, hold speed, and decrease speed.

\subsubsection{State}
The AI agent assessed the state using a top-down view of the virtual environment surrounding the passenger car (Figure \ref{fig:rlinput}). We used 3 successive video frames - a 3x64x64px grayscale image series - as the state input, $S$, to the deep learning network. With this image series, the AI agent could see the road, the position of both cars, and orb representations of objects at the side of the road. The luminance of these orb representations were based on their object category as targets or non-targets. When the car was near these orbs, it elicited a reinforcement signal, as described in the next section.

\subsubsection{Reward}
We rewarded the agent as long as the passenger car followed the lead car within prescribed bounds of distance. The agent received a positive reinforcement for staying within the bounds and a negative reinforcement when it violated these distance bounds (+1 and -10 reinforcement respectively). To include physiological signals into the reinforcement, the AI agent received an additional reward (or punishment) based on the neurophysiological response evoked by image objects within the visual distance of the passenger car; an object classified by the hBCI + TAG system as a target object yielded a reward while those classified as non-target objects yielded a penalty. We balanced the magnitude of reward and of penalty according to the prevalence of targets and non-targets in the environment (+3 and -1 reinforcement respectively). Some objects are misclassified by hBCI and therefore yield the wrong reinforcement (false positives and false negatives). In the false positive condition, an object with an orb representation with luminance corresponding to a non-target would yield positive reinforcement. In the false negative condition, an object with an orb representation with luminance corresponding to a target would yield negative reinforcement. For each subject, we chose to use the classification threshold which maximized the F1 score of that subject. The F1 score is a commonly used metric for determining a classifier's accuracy by using the harmonic mean of the precision and recall \cite{powers:evaluation}. The immediate reward, $R_{t+1}$, was the sum of all reinforcement values that are accumulated in the current rendered frame.

\begin{equation}
r_1 = \begin{cases} +1, & \mbox{if } 5< d < 60 \\  -10, & \mbox{else}\end{cases}
\end{equation}
\begin{equation}
r_2 = \begin{cases} 
+3, & \mbox{if } \text{WVD}_a \land \omega_a < \text{TPR}\\
-1, & \mbox{if } \text{WVD}_a \land \omega_a > \text{TPR}\\ 
0, & \mbox{else}
\end{cases}
\end{equation}
\begin{equation}
r_3 = \begin{cases} 
-1, & \mbox{if } \text{WVD}_{b} \land \omega_b > FPR\\ 
+3, & \mbox{if } \text{WVD}_{b} \land \omega_b < FPR\\
0, & \mbox{else}
\end{cases}
\end{equation}
\begin{equation}
R_{t+1} = f(w_1r_1 + w_2r_2 + w_3r_3)
\end{equation}

Where d is the distance between the passenger car and the lead car (at distances between 5 and 60, the AI was at a "safe distance" from the lead car); $\text{WVD}_{a}$ and $\text{WVD}_{b}$ are True when the car is within visual range of a target and a nontarget object respectively and False otherwise; TPR is the true positive rate and FPR is the false positive rate of the hBCI+TAG system derived from the subject; and $\omega$ is a uniformly distributed random number between 0 and 1 chosen at each incidence of WVD being True. In future instantiations, the different reinforcement schemes ($r_1, ... ,r_n$) could be weighted differently and the final value could be used in a function such as a sigmoid in order to squash the reinforcement value limits. In this study we weighted the reinforcement schemes equally at $w_1=w_2=w_3=1$ and did not use a squashing function.

\begin{figure*}
	\centering
	\begin{subfigure}[b]{0.4\textwidth}
		\centering
		\includegraphics[width=\textwidth]{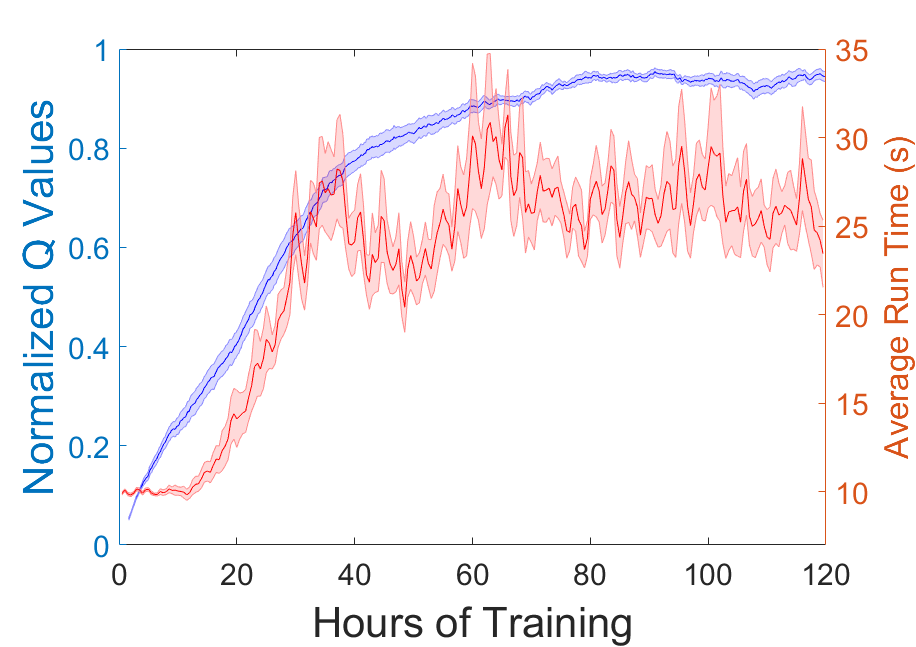}
		\caption[]%
		{{\small All Subjects Training Evolution}}    
		\label{fig:Q_Run}
	\end{subfigure}
	\begin{subfigure}[b]{0.4\textwidth}  
		\centering 
		\includegraphics[width=\textwidth]{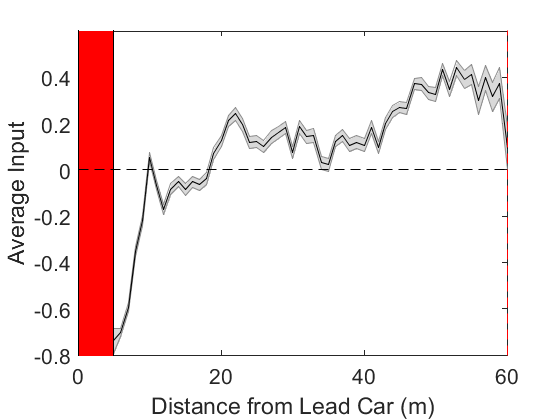}
		\caption[]%
		{{\small AI Agent Brake Behavior}}    
		\label{fig:brake_behavior}
	\end{subfigure}
	\caption
	{\small Training of the hBCI AI agent. a) This figure shows the normalized Q value averaged across subjects and the run time averaged across subjects through the training duration. The Q value evolution over training time suggests that all subjects have converged to a relatively stable Q value which indicates that training has plateaued. This observation is echoed by the driving performance which shows an increase of run time to a relatively stable average of around 25 seconds. The shaded area shows the standard error across subjects. b) This figure shows the learned brake behavior of the resulting AI agent. The areas that are marked in red indicate distances from the lead car which resulted in a game over and negative reinforcement. As expected, the car learns to avoid crashes by braking when it is too close and to avoid lagging too far by accelerating.} 
	\label{fig:mean and std of nets}
\end{figure*}

\begin{figure*}
	\centering
	\begin{subfigure}[b]{0.4\textwidth}  
		\centering 
		\includegraphics[width=\textwidth]{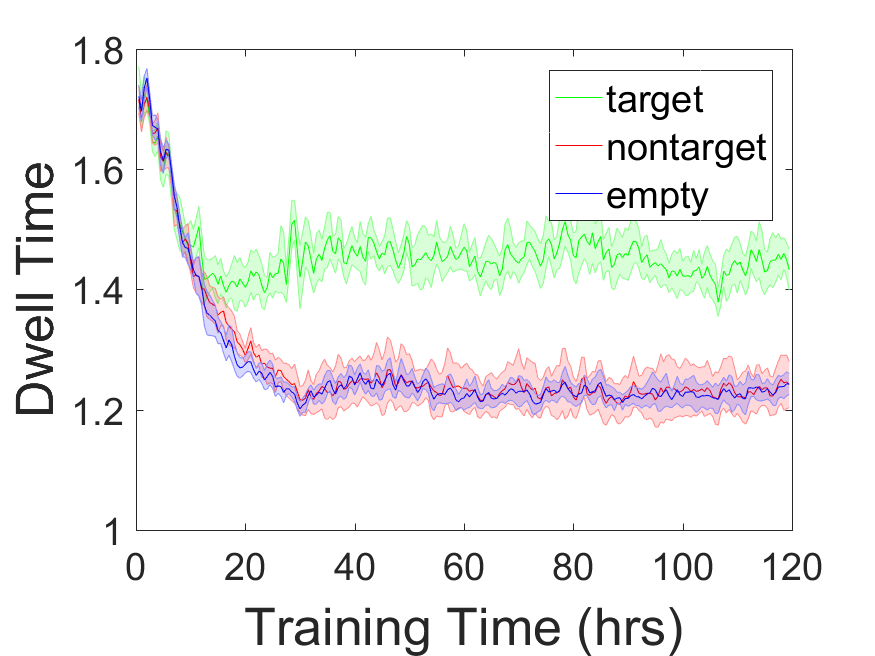}
		\caption[]%
		{{\small All Subjects Dwell Times}}    
		\label{fig:dwelltime}
	\end{subfigure}
	\begin{subfigure}[b]{0.4\textwidth}  
		\centering 
		\includegraphics[width=\textwidth]{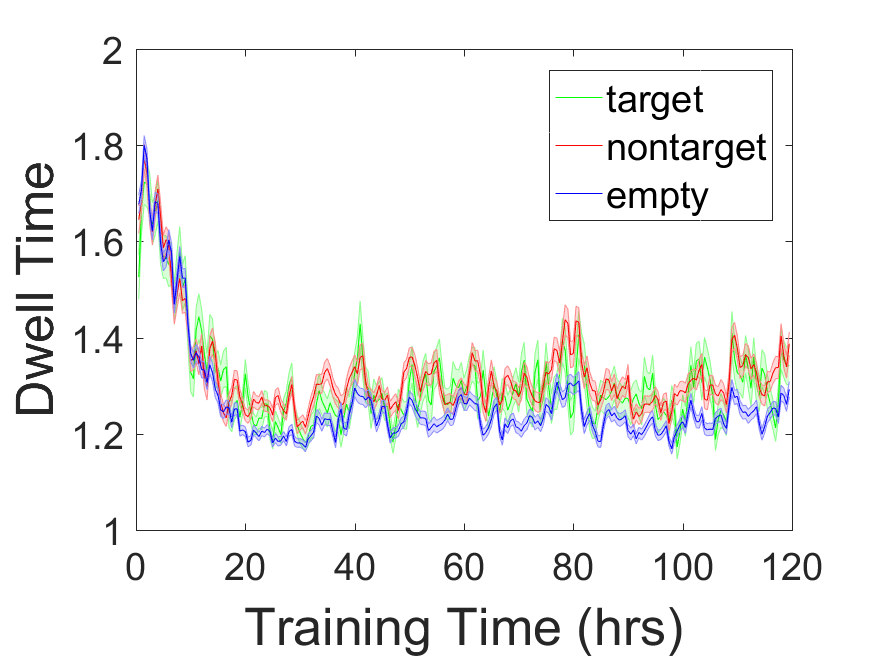}
		\caption[]%
		{{\small Control Subject Dwell Times}}    
		\label{fig:subjectcontrol}
	\end{subfigure}
	\caption
	{\small Results for hBCI Deep Learning Agent. a) Average dwell time between targets and non-targets show approximately 20\% increase in dwell time between targets and non-targets across subjects. The shaded area in the graph represents the standard error across subjects. b) A control subject was used to see how the AI agent behaved when a hBCI+TAG that outputted random classification values was used. The results show that there is very little separation of dwell times between targets, non-targets, and empty halls.}
		\label{fig:mean and std of nets}
\end{figure*}
\begin{figure*}
	\centering  
	\includegraphics[width=0.7\textwidth]{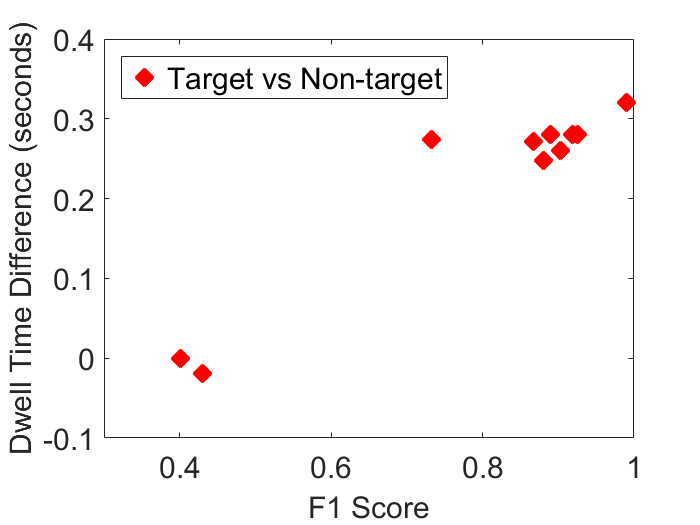}
	\caption[ The average and standard deviation of critical parameters ]
	{\small Comparing F1 score with the difference in dwell time shows that subjects with lower F1 score in the hBCI+TAG have a smaller separation of dwell times between targets and non-targets.} 
	\label{fig:dtvhp}
\end{figure*}

\section{Results}
We report results for 10 subjects; eight subjects showed a learned behavior based on their hBCI+TAG preferences while two subject did not show preference learning due to low SNR of the hBCI+TAG classification. Individual subject performance and Q-learning convergence can be found in the supplementary materials. Data from all subjects show a converging growth in Q-values during AI training, indicating that the AI agent is converging to a relatively stable policy for driving. (Figure \ref{fig:Q_Run}). This stable policy leads to an increase in the total run time of the autonomous vehicle in each independent driving session across all subjects. The AI agent learns the expected behavior, of braking when the passenger car gets too close to the lead car and accelerating when lagging too far behind, in order to stay within the reinforcement bounds. (Figure \ref{fig:brake_behavior})With the integration of hBCI reinforcement into the reward functions, this policy is also able to tune the behavior of the AI agent to each individual's subjective interest -- indicating whether they viewed each object as a target or non-target. One key metric of successful learning is the dwell time for each object type, which is the number of seconds that the passenger car stays within visual distance of an object.  Results show that the AI agent is able to differentiate between targets and non-targets, learning to keep the targets within view for a longer period of time (Figure \ref{fig:dwelltime}). As a control, we set the true positive rate and false positive rate to 0.5 to simulate a subject with an hBCI+TAG output that is random and observed that this control subject did not have significantly different dwell times between targets, nontargets, and empty halls (Figure \ref{fig:subjectcontrol}). As expected, the  success of the system in spending more time dwelling on targets (relative to non-targets or empty halls) depends on the F1 score of the hBCI+TAG classifier (Figure \ref{fig:dtvhp}). Specifically, we find that higher classification accuracy yields larger differences in dwell time between targets and non-targets. 

\begin{center}
	\captionof{table}{Subject hBCI+TAG Results} \label{tab:title}
	\begin{tabular}{ | c || c | c | c | }
		\hline
		Subject	& TPR	& FPR	& F1 Score	\\ \hline
		1		&0.8343	&0.0125	&0.8896		\\ \hline
		2		&0.9823	&0.9495	&0.4306		\\ \hline
		3		&1.0000	&0.0063	&0.9901		\\ \hline
		4		&0.8745	&0.0115	&0.9182		\\ \hline
		5		&0.8454	&0.0077	&0.9036		\\ \hline
		6		&0.8783	&0.0074	&0.9248		\\ \hline
		7		&0.8257	&0.0177	&0.8802		\\ \hline
		8		&1.0000	&0.9905	&0.4008		\\ \hline
		9		&0.7793	&0.0070	&0.8668		\\ \hline
		10		&0.6250	&0.0269	&0.7324		\\ \hline
	\end{tabular} 
\end{center}

\section{Discussion}

In this paper, we present a three-tiered machine learning approach (Figure ~\ref{fig:ml_fig}) for decoding neural and ocular signals reflecting a human passenger's level of interest and then using these subjective signals to favorably impact the driving strategy for an AI controlled vehicle. In our experiments, a favorable driving strategy is one that maintains a safe distance from a lead vehicle, slows the vehicle when objects of specific interest to the passenger are encountered during the drive, and ignores objects that do not interest the passenger. The prime novelty of our approach is that the human-machine interaction that communicates passenger preferences to the AI agent is implicit and via the hBCI -- i.e. the passenger does not need to communicate their preferences overtly, with say a press of a button, but instead preferences are inferred via decoded neural-ocular activity.  A second novel element of our approach is that we use semi-supervised CV-based learning to increase the prevalence of the reinforcement signals while also mitigating the relatively low signal to noise ratio (SNR) of the human neurophysiological data -- only a few evoked neural-ocular responses are needed to generate a model of the passenger's preferences. We show that in using a double-deepQ reinforcement architecture, we converge to a stable driving policy which increases the average run time of the autonomous vehicle. Additionally, in 8 out of 10 subjects (due to hBCI+TAG performance), the AI adapts to a driving strategy that significantly increases the time that the passengers can gaze at objects that are consistent with their individual interest. 

This approach can be used to tune driving behavior of an autonomous vehicle to individual preferences in several other scenarios. For example, reinforcement can be designed to optimize on other aspects of human preferences, such as whether the ride is ``comfortable" for the individual.  Here, ``comfortable" is a subjective metric that is specific to a particular human passenger and might be observed via changes in arousal, stress level, and emotional valence, amongst other physiological and cognitive factors. The importance of an AI agent recognizing and acting upon, or even predicting human preferences, is important not only as an interface but ultimately because it might be crucial for development of a "trusted relationship" between the human and machine akin to the emotional intelligence required for harmonious social interactions between humans \cite{hoff:trust,lewis:trust,saproo:cortically}. 

Currently, the system we describe utilizes EEG, pupil dilation, and eye position data as physiological signals used to train the classifier to distinguish targets from non-targets. Future investigations are needed to determine additional physiological and behavioral signals that can be fused in the hBCI to infer cognitive and emotional state. Specifically, for real-world applications where scalp EEG is not practical, using unobtrusive sensing modalities such as video to track facial micro-expressions as well as electrodermal activity (EDA) and Heart Rate Variability (HRV) might provide the necessary precision and recall to train reinforcement learning agents. 

Our current system is not currently closed-loop; it does not obtain new training data from the environment during the reinforcement process. Future work would entail closing the loop to modify AI behavior while new data is being continuously introduced into the system. We hypothesize that after the AI agent has learned to slow down when the car approaches non-targets, the hBCI will yield more accurate inferences because the subject is more likely to attend to the targets in a future run. This new more accurate data can be propagated in the TAG module to train the AI agent again and further improve the differentiation between targets and non-targets.


%

\section*{Acknowledgment}

	The work was partially funded by the Army Research Laboratory under Cooperative agreement number W911NF-10-2-0022. This research was partially supported by BRAIQ, Inc.

\ifCLASSOPTIONcaptionsoff
  \newpage
\fi



%
\bibliographystyle{IEEEtran}
\bibliography{Towards_Personalized_AI}



\clearpage
\appendices
\counterwithin{figure}{section}
\section{Dwell Time Figures}
\noindent\begin{minipage}{\textwidth}
	\begin{minipage}{.33\textwidth}
		\centering
		\centering
		\includegraphics[width=\textwidth]{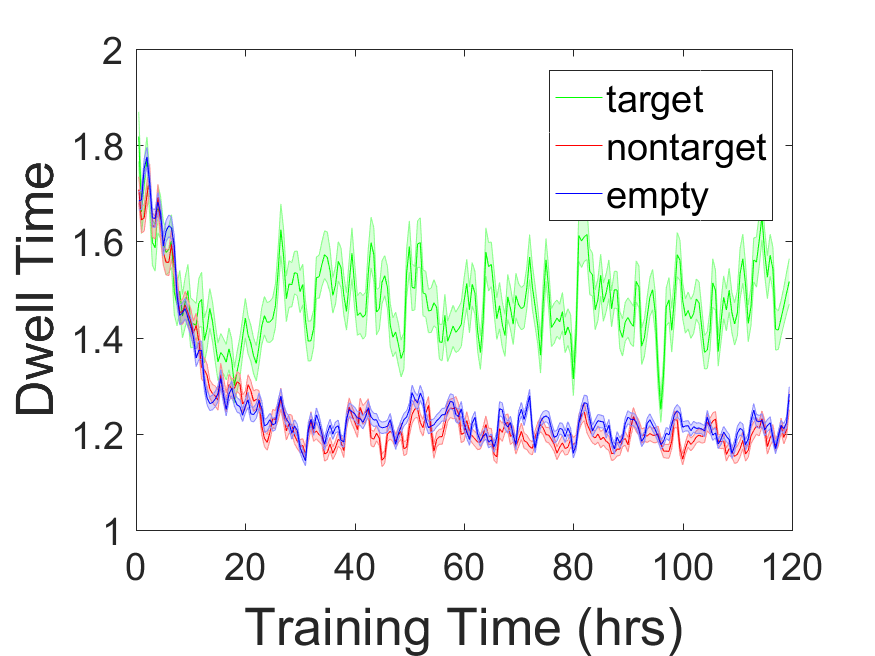}
		\captionof{subfigure}{Dwell Time for Subject 1}
		\label{fig:jjj}
	\end{minipage}%
	\begin{minipage}{.33\textwidth}
		\centering
		\centering
		\includegraphics[width=\textwidth]{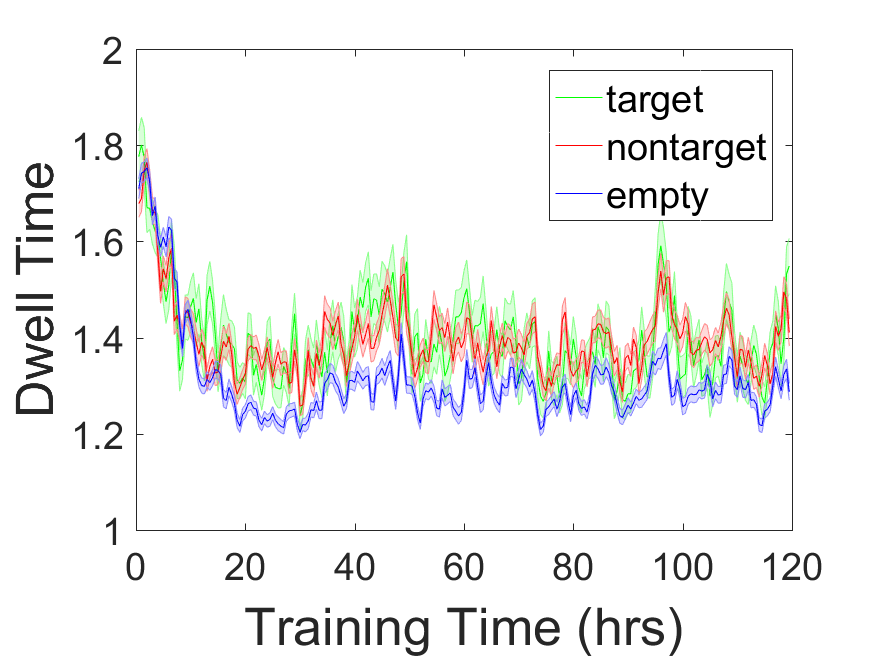}
		\captionof{subfigure}{Dwell Time for Subject 2}
		\label{fig:jjj}
	\end{minipage}%
	\vspace{.01\vsize}
	\begin{minipage}{.33\textwidth}
		\centering
		\centering
		\includegraphics[width=\textwidth]{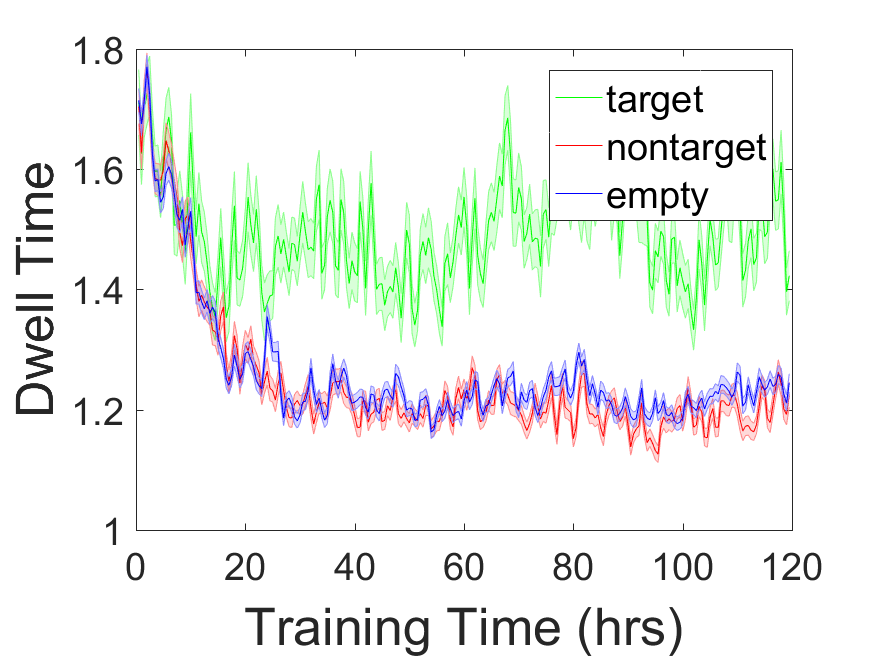}
		\captionof{subfigure}{Dwell Time for Subject 3}
		\label{fig:jjj}
	\end{minipage}%
	\vspace{.01\vsize}
	\begin{minipage}{.33\textwidth}
		\centering
		\centering
		\includegraphics[width=\textwidth]{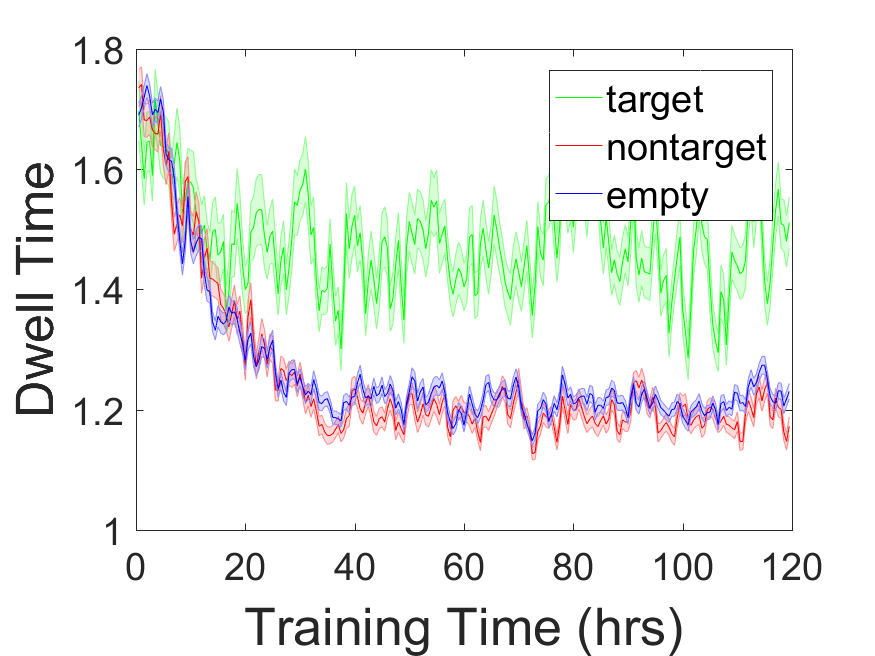}
		\captionof{subfigure}{Dwell Time for Subject 4}
		\label{fig:jjj}
	\end{minipage}%
	\begin{minipage}{.33\textwidth}
		\centering
		\centering
		\includegraphics[width=\textwidth]{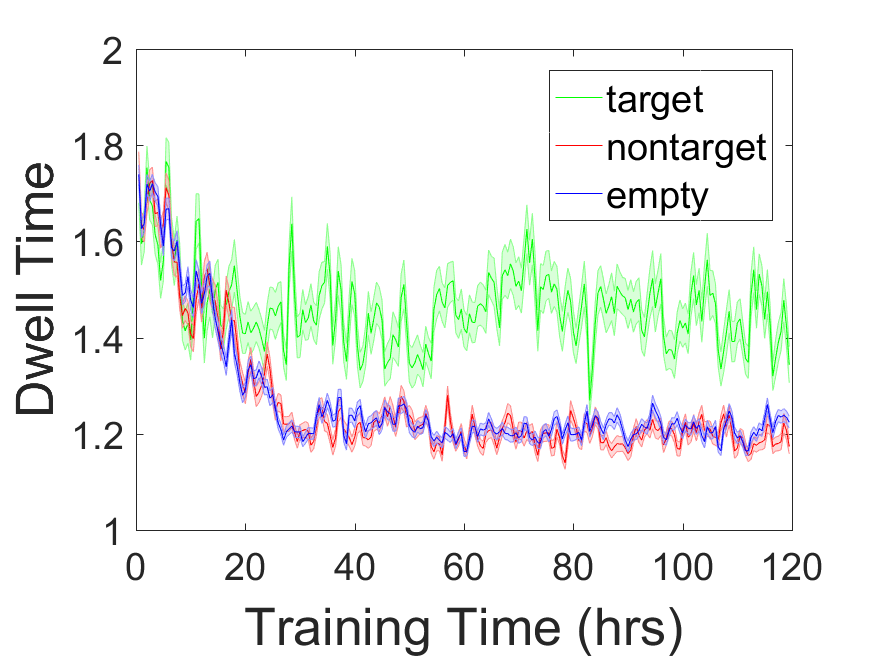}
		\captionof{subfigure}{Dwell Time for Subject 5}
		\label{fig:jjj}
	\end{minipage}%
	\begin{minipage}{.33\textwidth}
		\centering
		\centering
		\includegraphics[width=\textwidth]{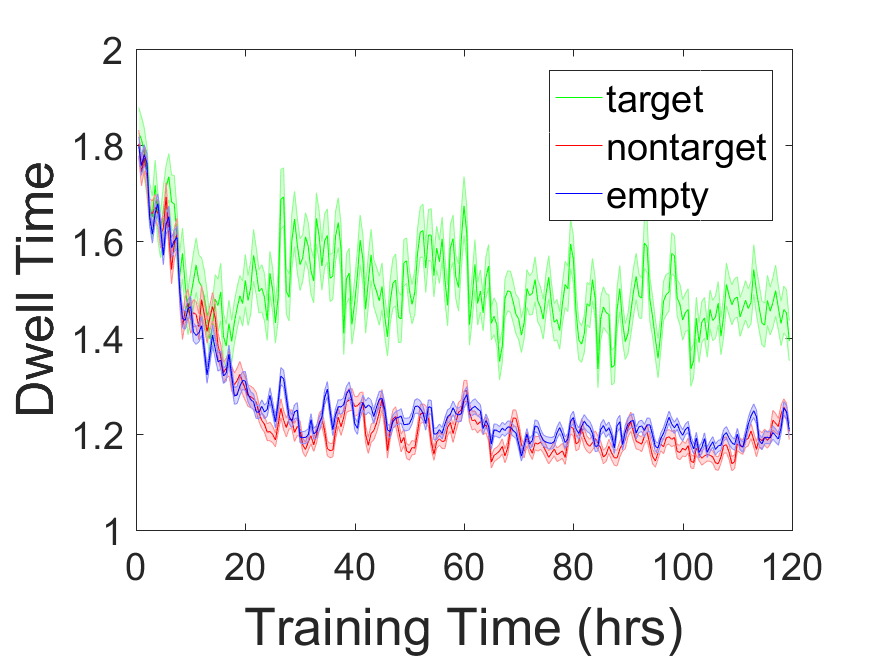}
		\captionof{subfigure}{Dwell Time for Subject 6}
		\label{fig:jjj}
	\end{minipage}%
	\vspace{.01\vsize}
	\begin{minipage}{.33\textwidth}
		\centering
		\centering
		\includegraphics[width=\textwidth]{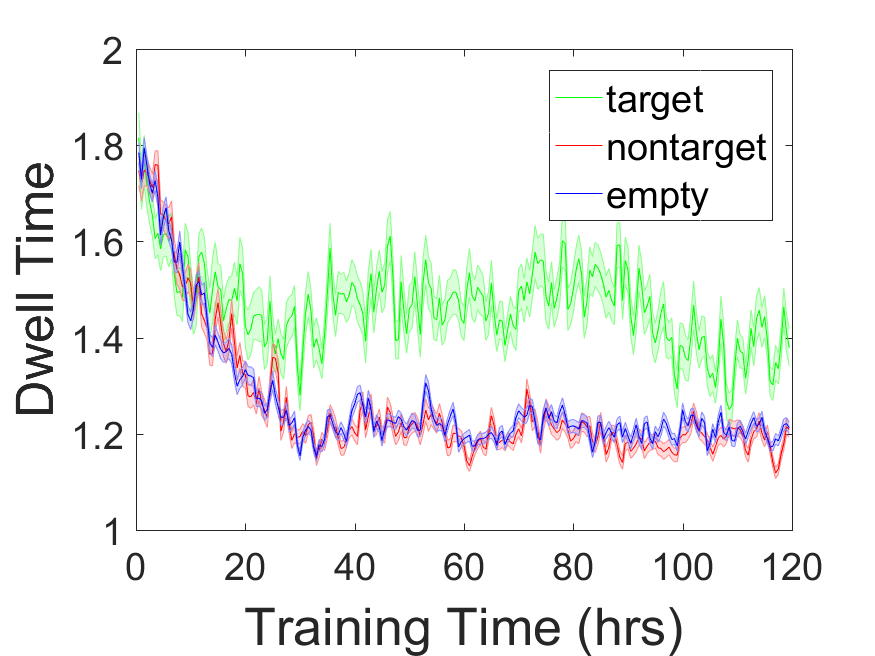}
		\captionof{subfigure}{Dwell Time for Subject 7}
		\label{fig:jjj}
	\end{minipage}%
	\begin{minipage}{.33\textwidth}
		\centering
		\centering
		\includegraphics[width=\textwidth]{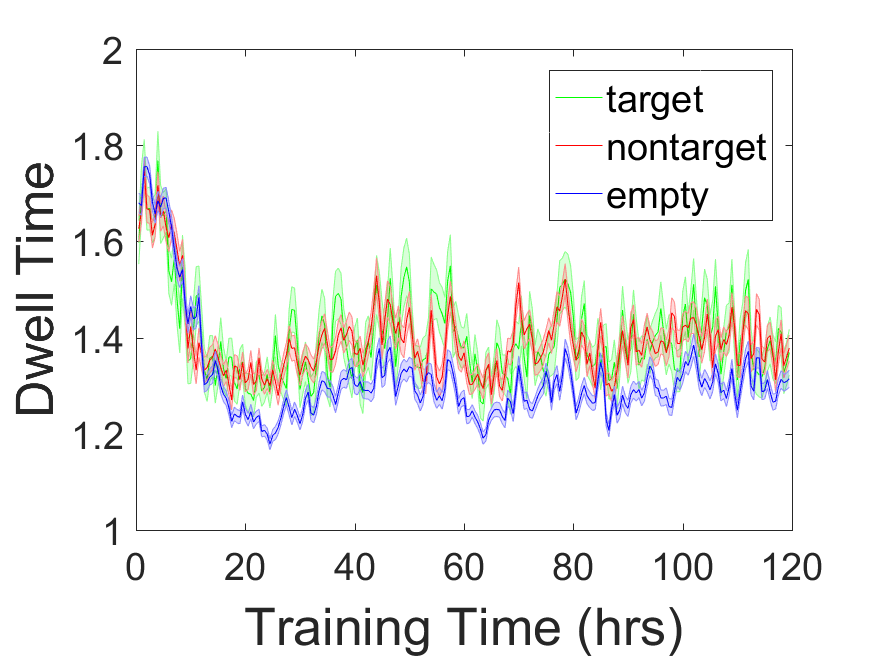}
		\captionof{subfigure}{Dwell Time for Subject 8}
		\label{fig:jjj}
	\end{minipage}%
	\begin{minipage}{.33\textwidth}
		\centering
		\centering
		\includegraphics[width=\textwidth]{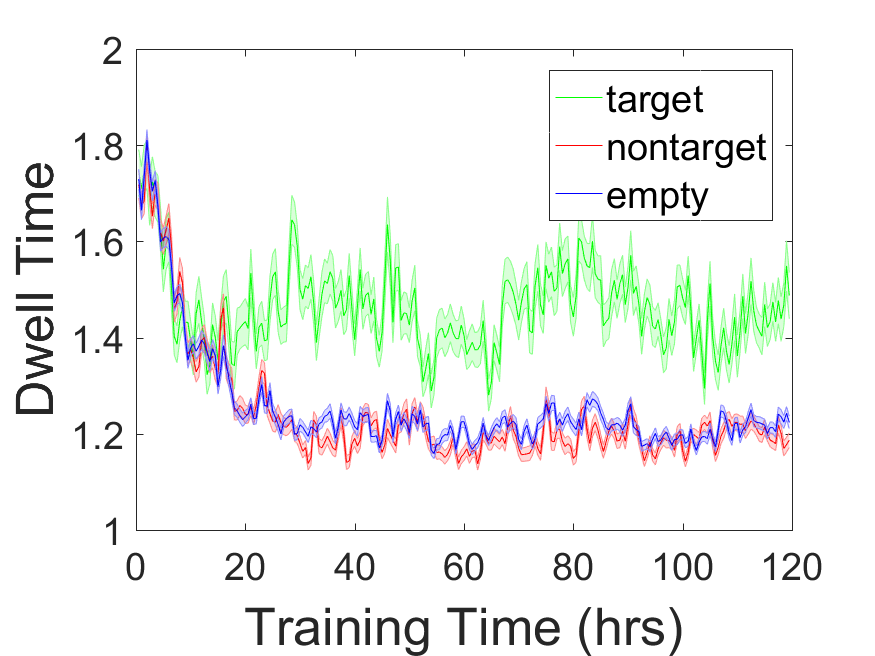}
		\captionof{subfigure}{Dwell Time for Subject 9}
		\label{fig:jjj}
	\end{minipage}%
	\vspace{.01\vsize}
	\begin{minipage}{.33\textwidth}
		\centering
		\centering
		\includegraphics[width=\textwidth]{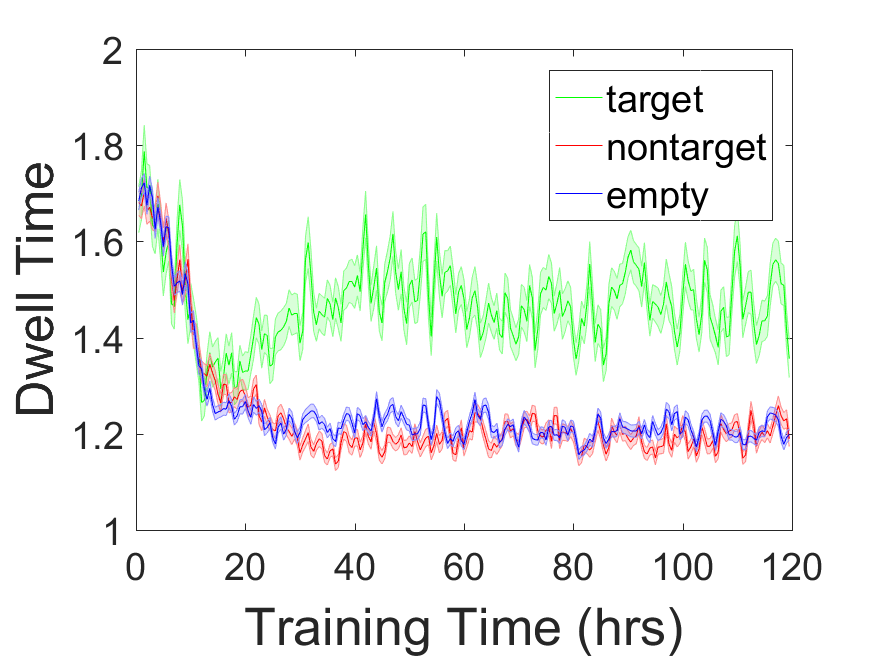}
		\captionof{subfigure}{Dwell Time for Subject 10}
		\label{fig:jjj}
	\end{minipage}%
\end{minipage}
\clearpage

\section{Run Time Figures}
\noindent\begin{minipage}{\textwidth}
	\begin{minipage}{.33\textwidth}
		\centering
		\centering
		\includegraphics[width=\textwidth]{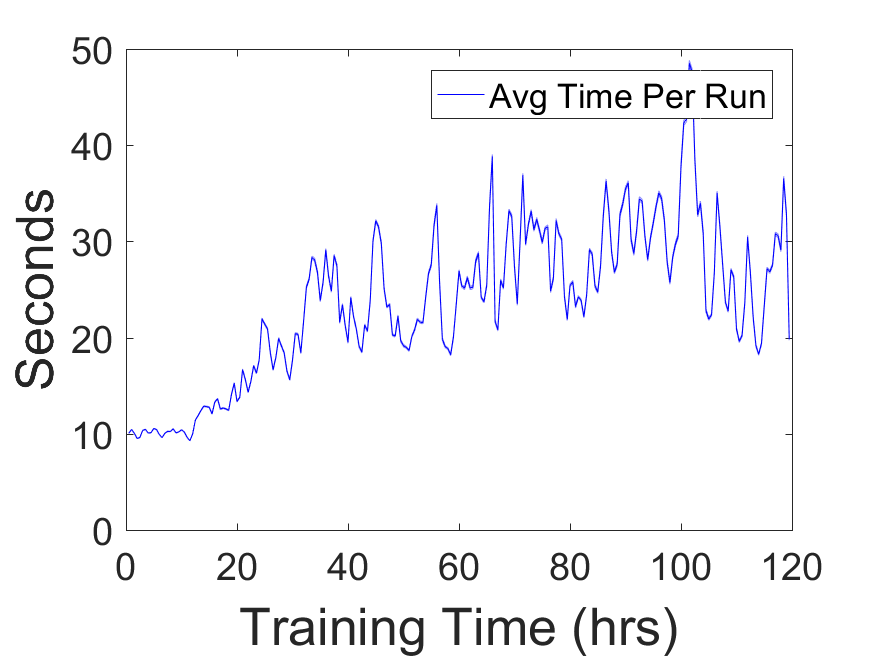}
		\captionof{subfigure}{Run Time for Subject 1}
		\label{fig:jjj}
	\end{minipage}%
	\begin{minipage}{.33\textwidth}
		\centering
		\centering
		\includegraphics[width=\textwidth]{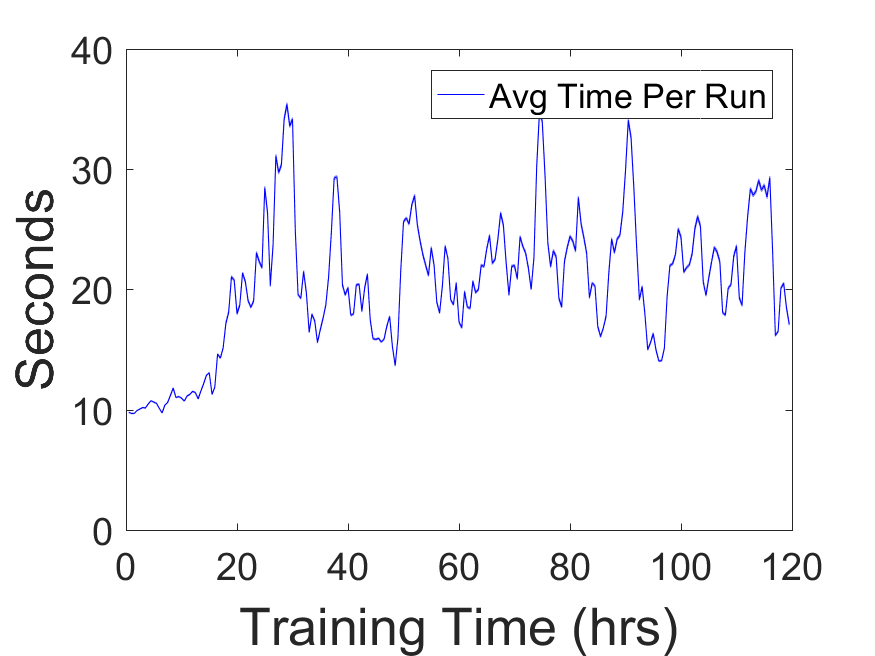}
		\captionof{subfigure}{Run Time for Subject 2}
		\label{fig:jjj}
	\end{minipage}%
	\vspace{.01\vsize}
	\begin{minipage}{.33\textwidth}
		\centering
		\centering
		\includegraphics[width=\textwidth]{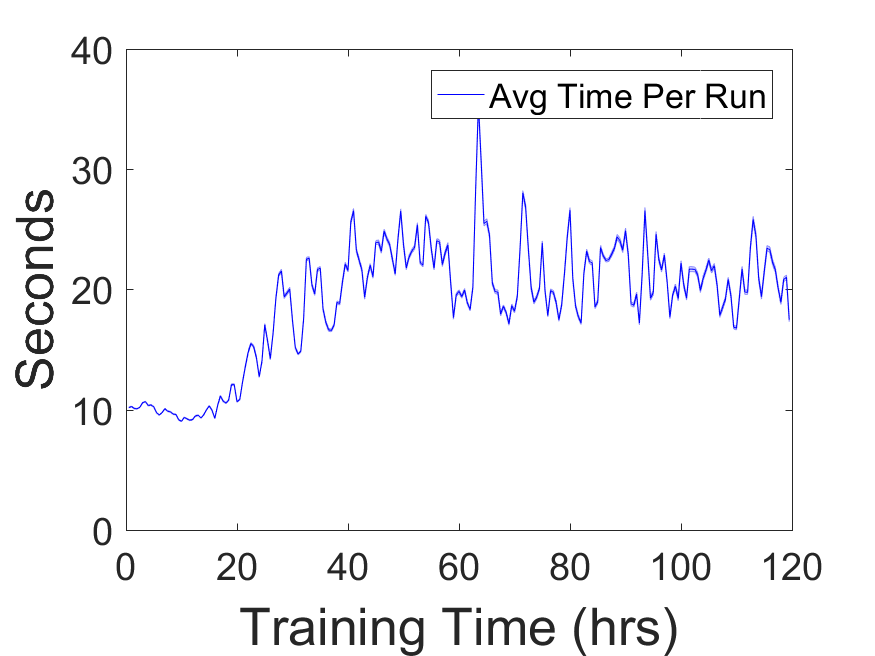}
		\captionof{subfigure}{Run Time for Subject 3}
		\label{fig:jjj}
	\end{minipage}%
	\vspace{.01\vsize}
	\begin{minipage}{.33\textwidth}
		\centering
		\centering
		\includegraphics[width=\textwidth]{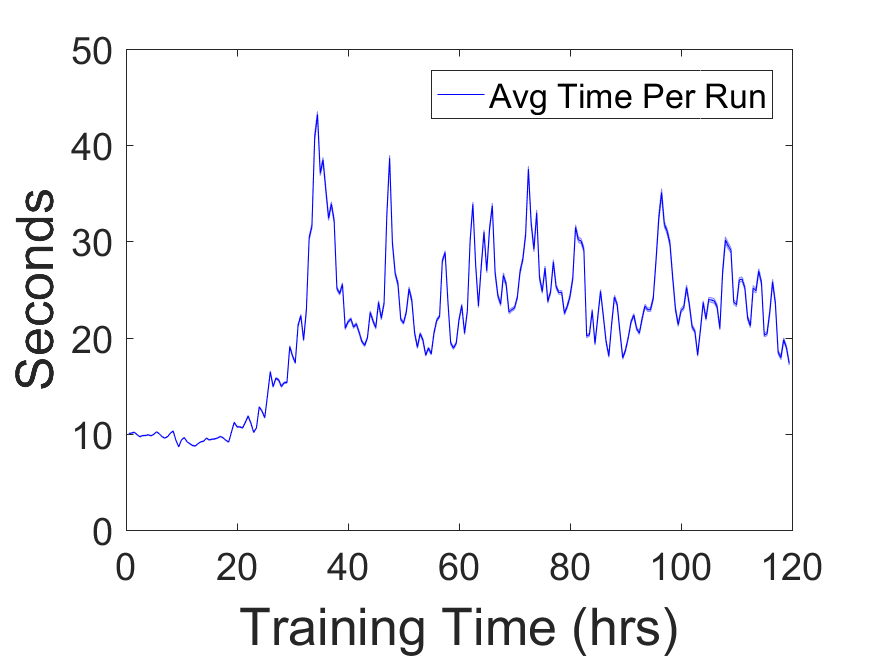}
		\captionof{subfigure}{Run Time for Subject 4}
		\label{fig:jjj}
	\end{minipage}%
	\begin{minipage}{.33\textwidth}
		\centering
		\centering
		\includegraphics[width=\textwidth]{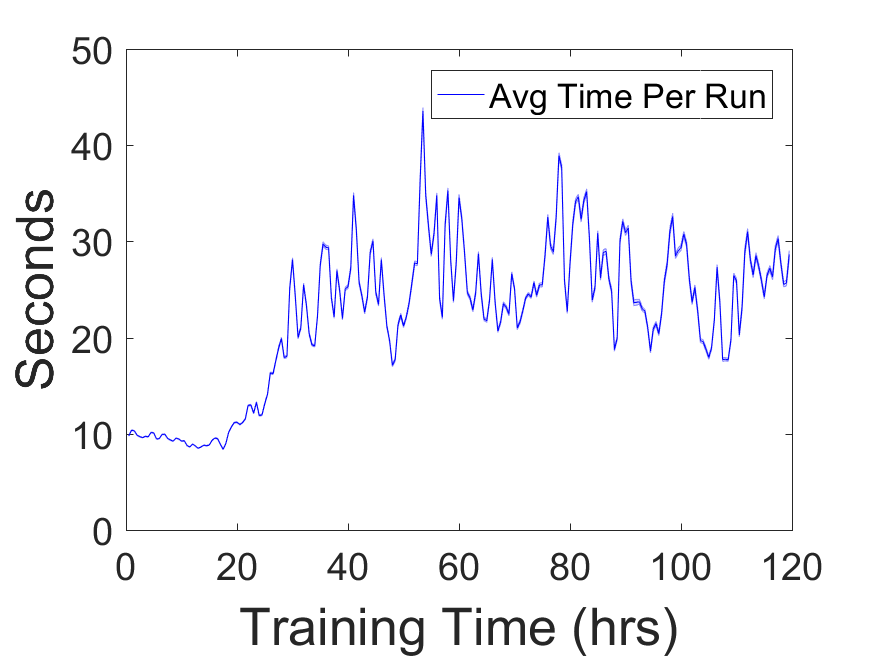}
		\captionof{subfigure}{Run Time for Subject 5}
		\label{fig:jjj}
	\end{minipage}%
	\begin{minipage}{.33\textwidth}
		\centering
		\centering
		\includegraphics[width=\textwidth]{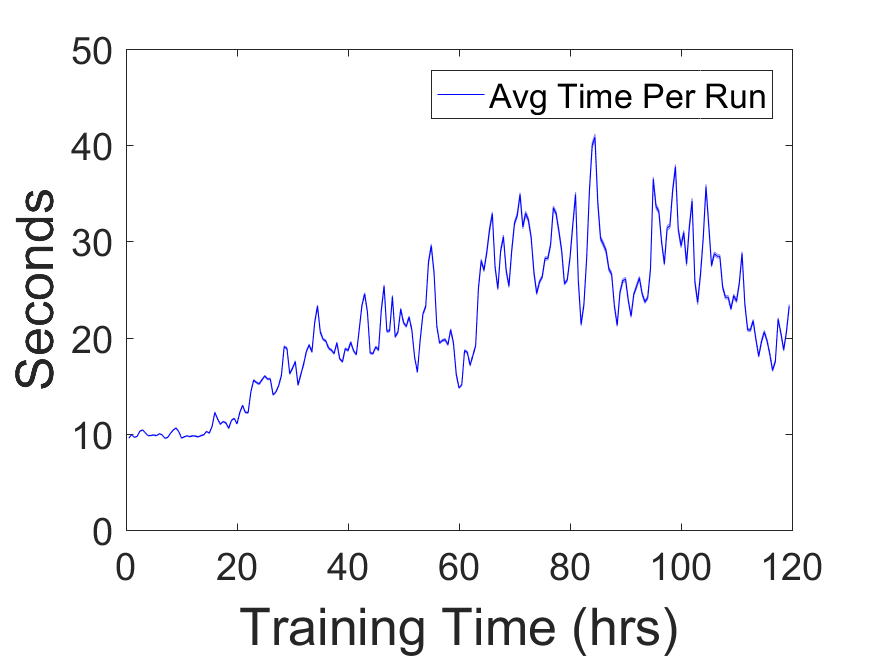}
		\captionof{subfigure}{Run Time for Subject 6}
		\label{fig:jjj}
	\end{minipage}%
	\vspace{.01\vsize}
	\begin{minipage}{.33\textwidth}
		\centering
		\centering
		\includegraphics[width=\textwidth]{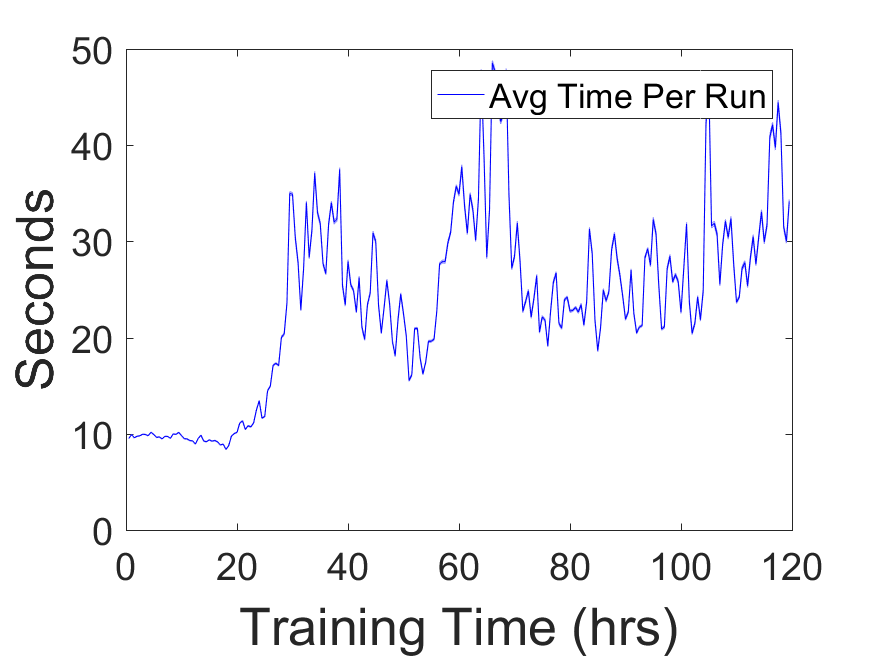}
		\captionof{subfigure}{Run Time for Subject 7}
		\label{fig:jjj}
	\end{minipage}%
	\begin{minipage}{.33\textwidth}
		\centering
		\centering
		\includegraphics[width=\textwidth]{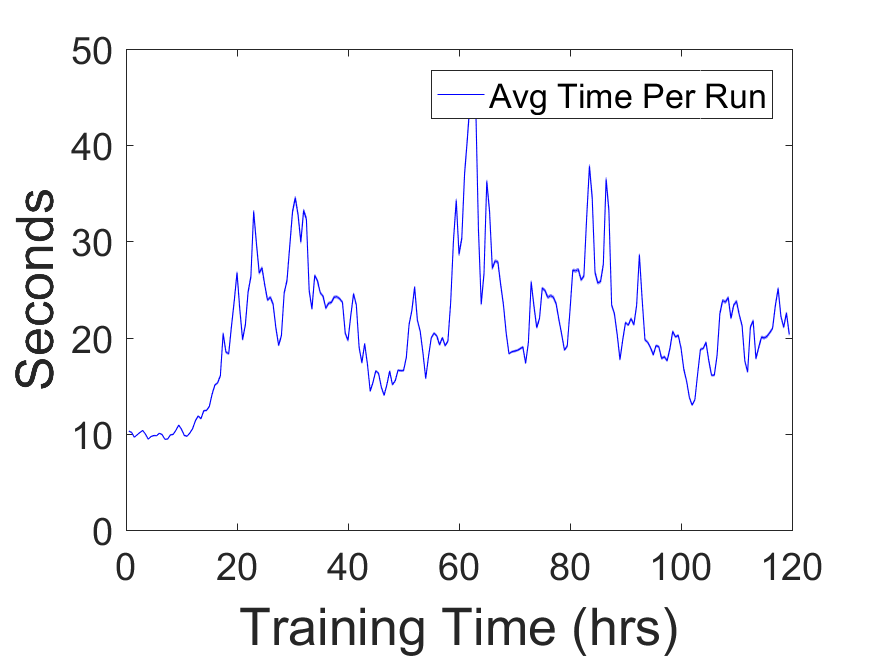}
		\captionof{subfigure}{Run Time for Subject 8}
		\label{fig:jjj}
	\end{minipage}%
	\begin{minipage}{.33\textwidth}
		\centering
		\centering
		\includegraphics[width=\textwidth]{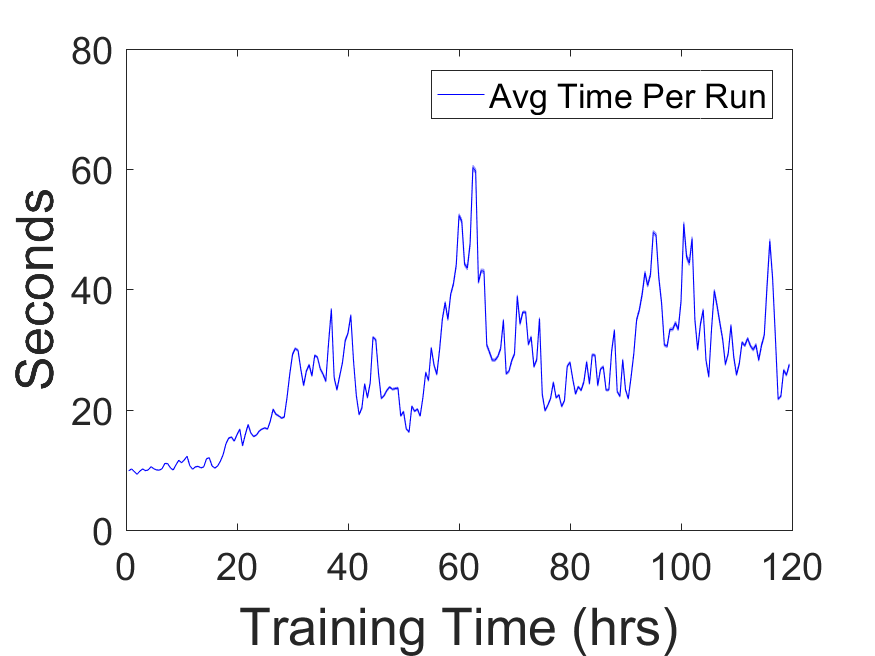}
		\captionof{subfigure}{Run Time for Subject 9}
		\label{fig:jjj}
	\end{minipage}%
	\vspace{.01\vsize}
	\begin{minipage}{.33\textwidth}
		\centering
		\centering
		\includegraphics[width=\textwidth]{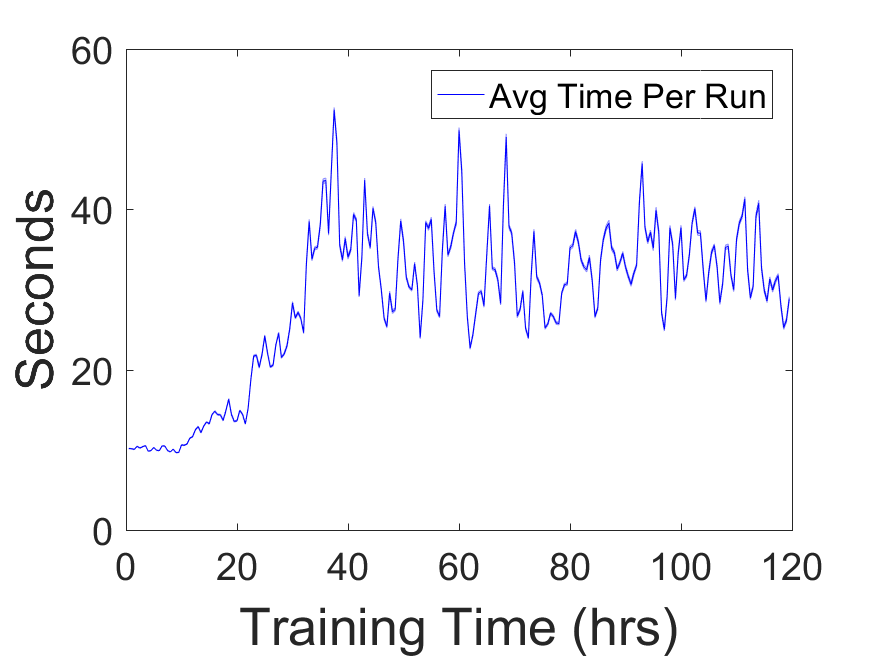}
		\captionof{subfigure}{Run Time for Subject 10}
		\label{fig:jjj}
	\end{minipage}%
\end{minipage}
\clearpage

\section{Q-Value Figures}
\noindent\begin{minipage}{\textwidth}
	\begin{minipage}{.33\textwidth}
		\centering
		\centering
		\includegraphics[width=\textwidth]{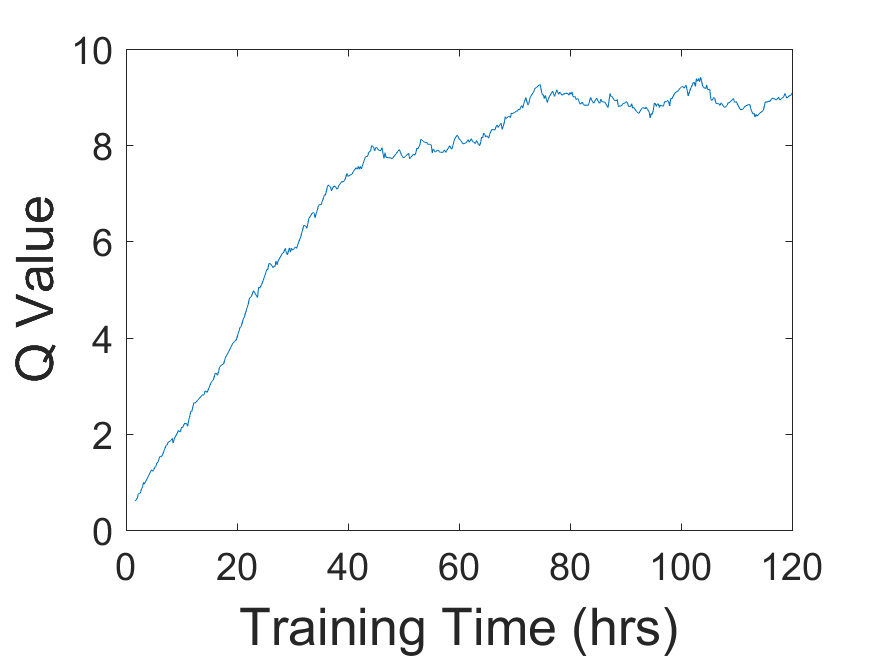}
		\captionof{subfigure}{Q Values for Subject 1}
		\label{fig:jjj}
	\end{minipage}%
	\begin{minipage}{.33\textwidth}
		\centering
		\centering
		\includegraphics[width=\textwidth]{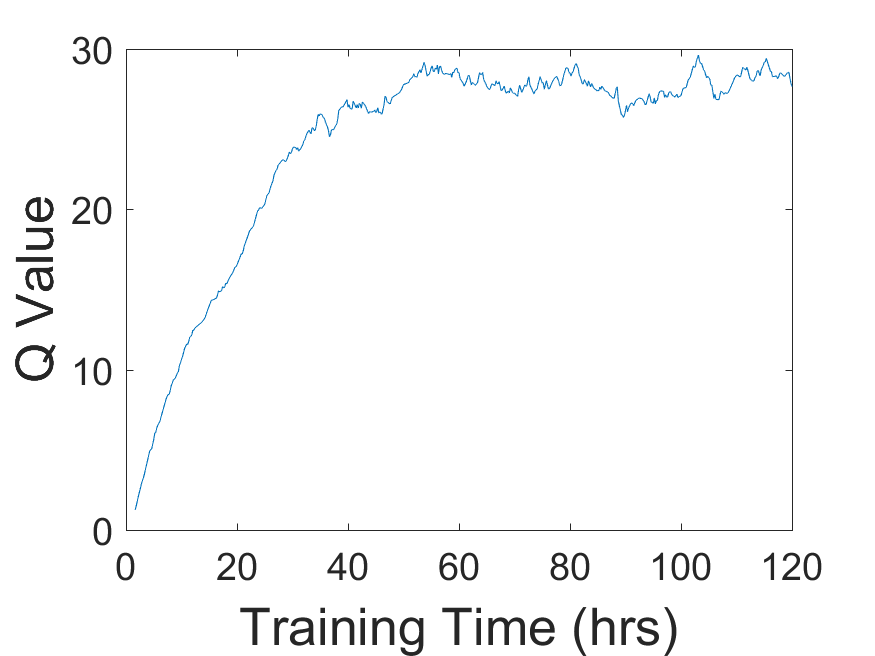}
		\captionof{subfigure}{Q Values for Subject 2}
		\label{fig:jjj}
	\end{minipage}%
	\vspace{.01\vsize}
	\begin{minipage}{.33\textwidth}
		\centering
		\centering
		\includegraphics[width=\textwidth]{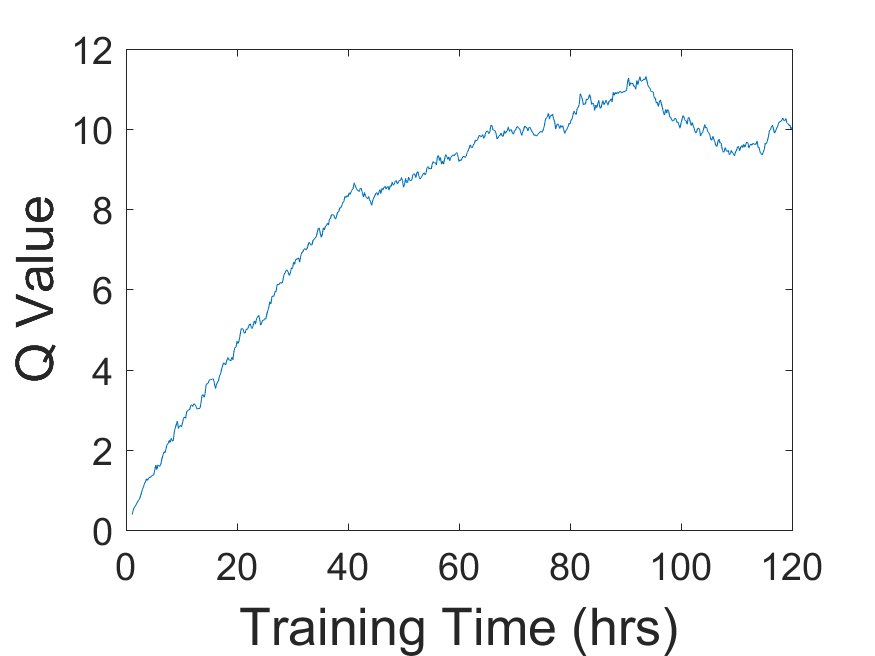}
		\captionof{subfigure}{Q Values for Subject 3}
		\label{fig:jjj}
	\end{minipage}%
	\vspace{.01\vsize}
	\begin{minipage}{.33\textwidth}
		\centering
		\centering
		\includegraphics[width=\textwidth]{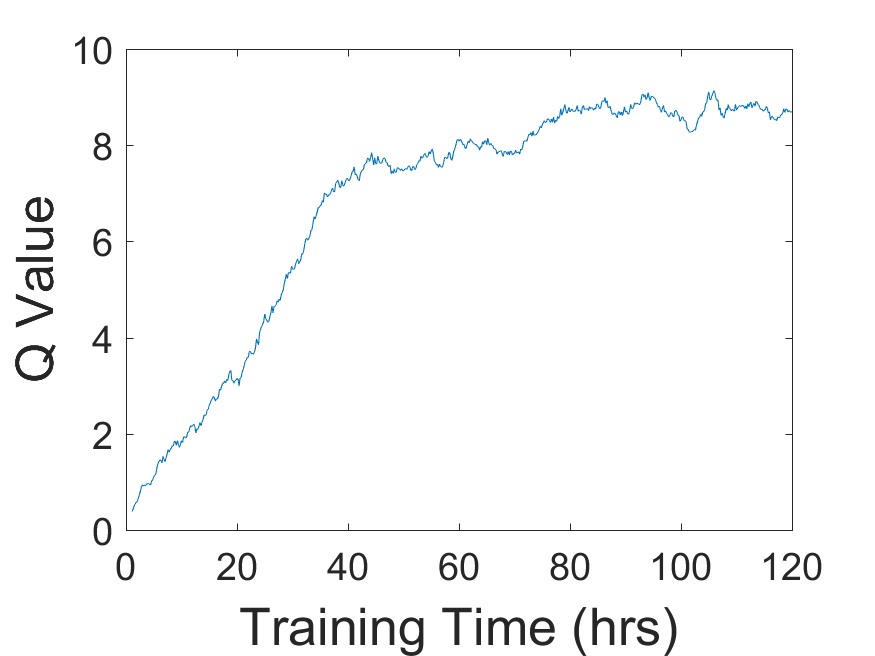}
		\captionof{subfigure}{Q Values for Subject 4}
		\label{fig:jjj}
	\end{minipage}%
	\begin{minipage}{.33\textwidth}
		\centering
		\centering
		\includegraphics[width=\textwidth]{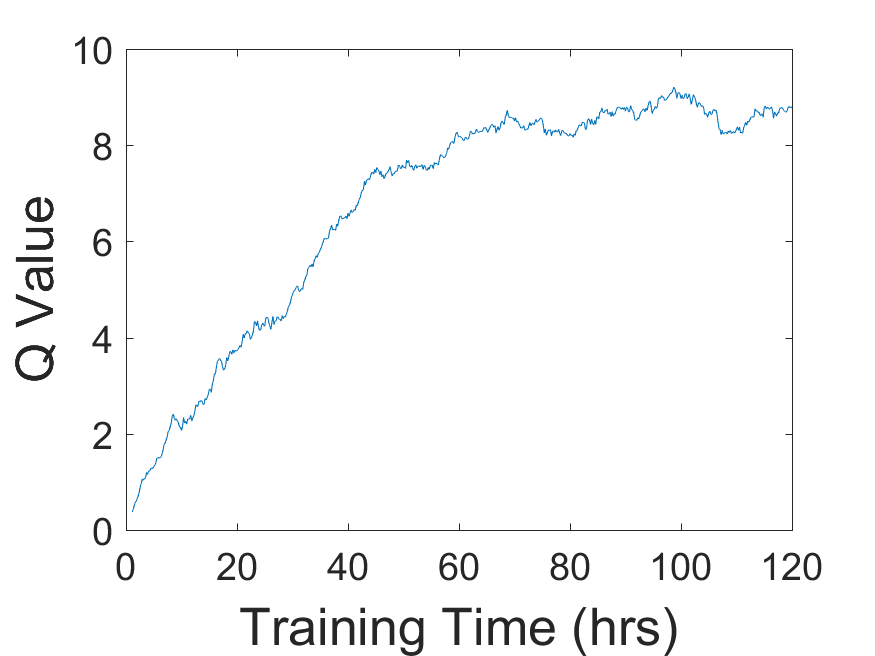}
		\captionof{subfigure}{Q Values for Subject 5}
		\label{fig:jjj}
	\end{minipage}%
	\begin{minipage}{.33\textwidth}
		\centering
		\centering
		\includegraphics[width=\textwidth]{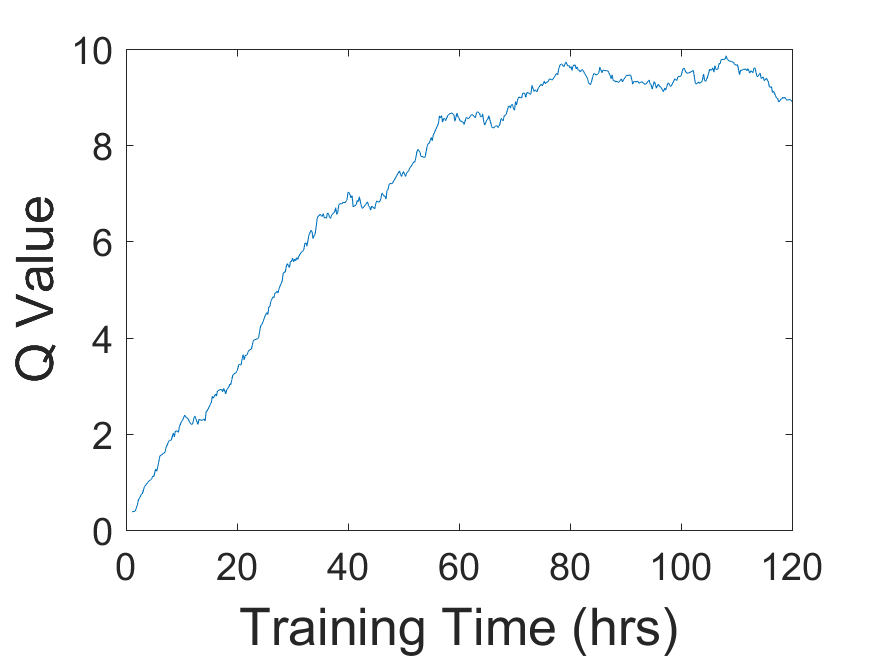}
		\captionof{subfigure}{Q Values for Subject 6}
		\label{fig:jjj}
	\end{minipage}%
	\vspace{.01\vsize}
	\begin{minipage}{.33\textwidth}
		\centering
		\centering
		\includegraphics[width=\textwidth]{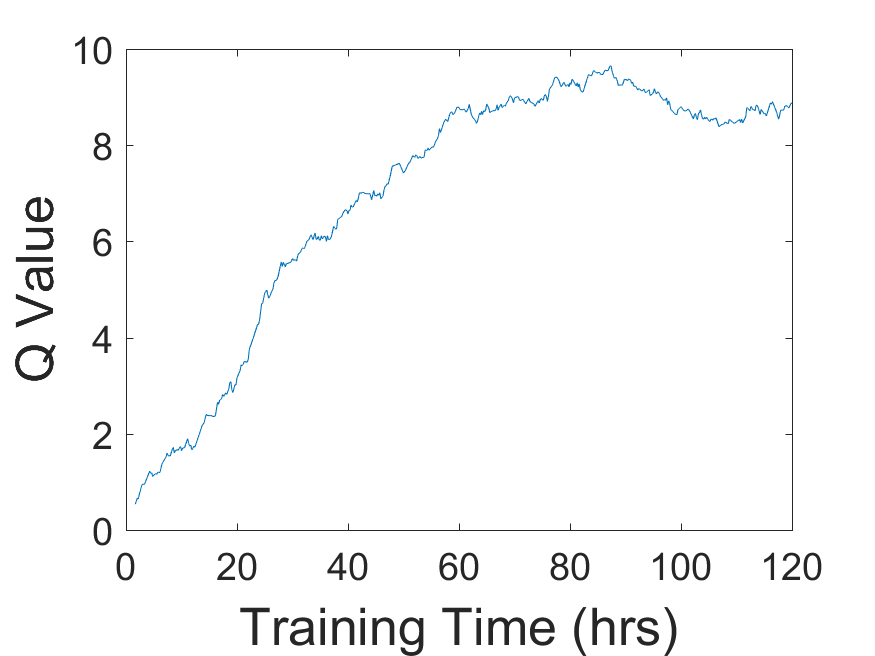}
		\captionof{subfigure}{Q Values for Subject 7}
		\label{fig:jjj}
	\end{minipage}%
	\begin{minipage}{.33\textwidth}
		\centering
		\centering
		\includegraphics[width=\textwidth]{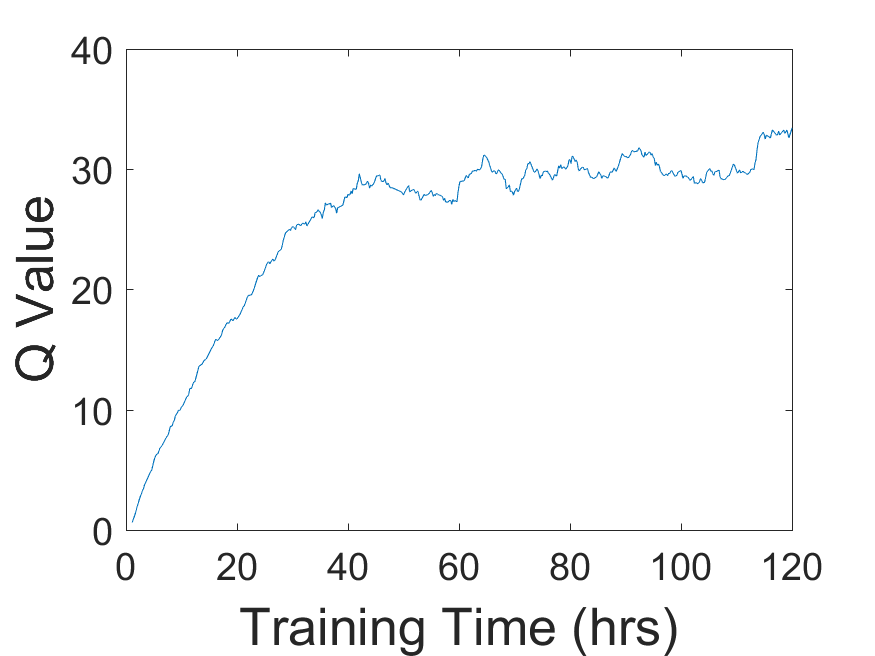}
		\captionof{subfigure}{Q Values for Subject 8}
		\label{fig:jjj}
	\end{minipage}%
	\begin{minipage}{.33\textwidth}
		\centering
		\centering
		\includegraphics[width=\textwidth]{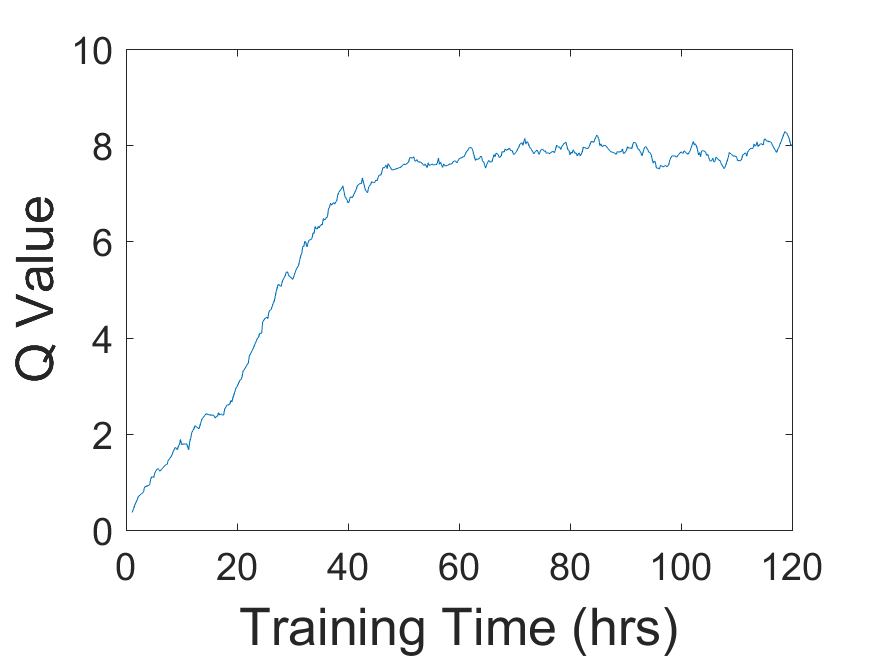}
		\captionof{subfigure}{Q Values for Subject 9}
		\label{fig:jjj}
	\end{minipage}%
	\vspace{.01\vsize}
	\begin{minipage}{.33\textwidth}
		\centering
		\centering
		\includegraphics[width=\textwidth]{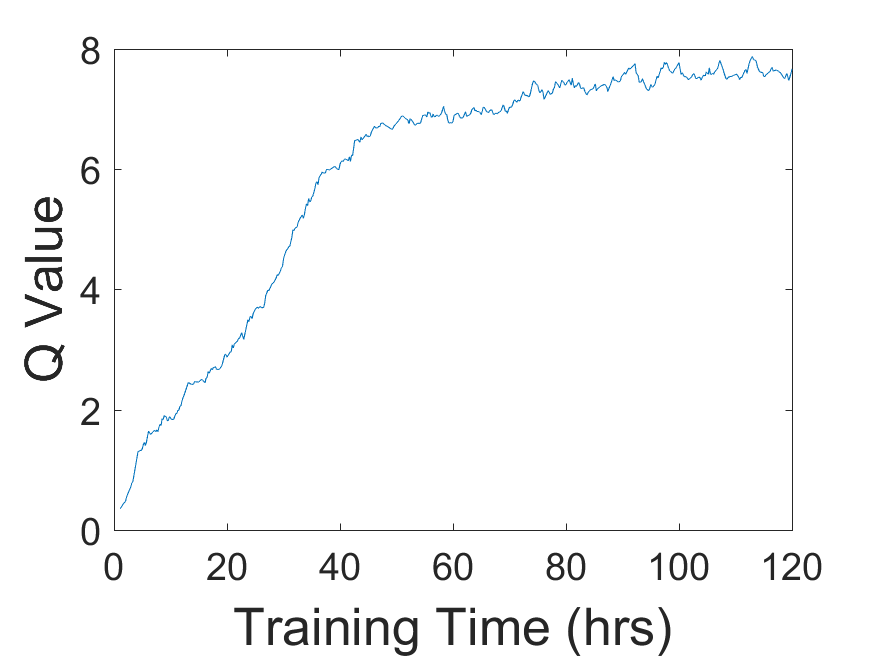}
		\captionof{subfigure}{Q Values for Subject 10}
		\label{fig:jjj}
	\end{minipage}%
\end{minipage}
\end{document}